\documentclass[%
 reprint,
 amsmath,amssymb,
 aps,
]{revtex4-2}

\usepackage{graphicx}
\usepackage{dcolumn}
\usepackage{bm}

\usepackage[final]{pdfcomment} 
\usepackage{accsupp}           
\usepackage{pgf}               

\newcommand\textopacity[2]{%
  \begin{pgfpicture}%
    \pgfsetfillopacity{#1}%
    \pgfpathmoveto{\pgfpointorigin}%
    \pgftext[base]{#2}%
  \end{pgfpicture}%
}

\makeatletter
\newcommand\textstack[3][;]{%
  \def\textstack@append##1{\expandafter\textstack@step##1#1\@eol}%
  \def\textstack@step##1#1##2\@eol{%
    \hbox{\textopacity{#2}{\ignorespaces ##1\unskip}}%
    \ifx\@eol##2\@eol\else
      \vskip-\baselineskip
      \textstack@step##2\@eol
    \fi}%
  \pdftooltip{%
    \BeginAccSupp{method=pdfstringdef,unicode,ActualText={#3}}%
      \vbox{\textstack@append{#3}}%
    \EndAccSupp{}%
  }{#3}%
}
\makeatother

\begin{document}

\preprint{APS/123-QED}


\author{Yoong S. Phang}
\author{Artem V. Talanov}
\affiliation{%
 Harvard University, Cambridge, MA, USA
}%

\collaboration{In collaboration with ChatGPT and Midjourney}

\date{\today}

\title{Engineered Graviton Condensates in a Room-Temperature Superconductor for a Unified Quantum Fibonacci Field Theory}

\begin{abstract}
In the quest to unify quantum mechanics with general relativity, the concept of gravitons as quantum carriers of gravitational force stands as a pivotal yet unproven hypothesis. This work pioneers a bold approach to graviton condensation via an innovative ambient superconductor, synthesized through a groundbreaking yet theoretically plausible physics process. Building on recent superconductivity breakthroughs, we introduce a novel ambient superconductor fabricated through the bombardment of a dilute Bose-Einstein condensate with high-energy gamma photons, under precisely controlled magnetic fields oscillating in Fibonacci sequence patterns. This process, designed to exploit quantum coherence at macroscopic scales, fosters conditions ripe for graviton condensation. Here we demonstrate the successful induction of a graviton condensate within our engineered superconductor, marking a significant stride towards integrating gravitational and electromagnetic forces at a quantum level. This discovery not only defies traditional views separating the four fundamental interactions but also lays down a tangible groundwork for quantum gravity. Our results challenge existing paradigms by indicating that graviton condensation, mediated through tailored superconducting states, can unlock new aspects of fundamental physics. This graviton condensate represents a critical step towards a unified field theory, bridging the gap between quantum mechanics and general relativity. This investigation not only extends the boundaries of theoretical physics but also hints at potential future technologies harnessing the quantum interplay of gravitational and electromagnetic forces. Through imaginative yet rigorous scientific inquiry, this study underscores the limitless potential of physics to explore and unify the natural world's most fundamental forces.
\end{abstract}

\maketitle

\section{\label{sec:level1}Introduction}

The pursuit of a unified theory that seamlessly integrates the gravitational force with the quantum realm has long stood as one of the most profound challenges in theoretical physics. The general theory of relativity, which elegantly describes gravity as the curvature of spacetime caused by mass, contrasts starkly with the principles of quantum mechanics, which governs the probabilistic behavior of particles at microscopic scales. This dichotomy has led to an intense search for a theory of quantum gravity, a framework that can reconcile these two pillars of modern physics.

Among the various candidates for mediating quantum gravitational interactions, gravitons—hypothetical massless spin-2 particles—offer a compelling pathway\cite{QuantumGravitationalInteractions2020,Liang2024}. In quantum field theory, gravitons are envisioned as the quanta of the gravitational field, analogous to photons in the electromagnetic field. However, the incredibly weak coupling of gravitons with matter, coupled with the non-renormalizable nature of gravity at quantum scales, has precluded their direct detection and the validation of theories involving them.

Recent advances in the field of superconductivity have opened new avenues for exploring quantum gravitational effects. Superconductors, materials that exhibit zero electrical resistance and the expulsion of magnetic fields when cooled below a critical temperature, embody quantum coherence on macroscopic scales. This property, intriguingly, offers a novel context in which gravitational and electromagnetic interactions could potentially converge in quantized form.

\subsection{Graviton Condensation and Ambient Superconductors}
The concept of particle condensates, particularly Bose-Einstein condensates (BECs), provides a framework for understanding how particles can occupy the same quantum state en masse, leading to macroscopic quantum phenomena. Inspired by this, we hypothesize the existence of a graviton condensate, a state in which gravitons collectively occupy a single quantum state, potentially enabling the direct quantization of gravitational fields. Achieving such a state, however, demands a medium where gravitons can interact strongly enough to condense, a requirement seemingly at odds with their inherently weak coupling nature.

\begin{figure*}
\includegraphics[width=\textwidth]{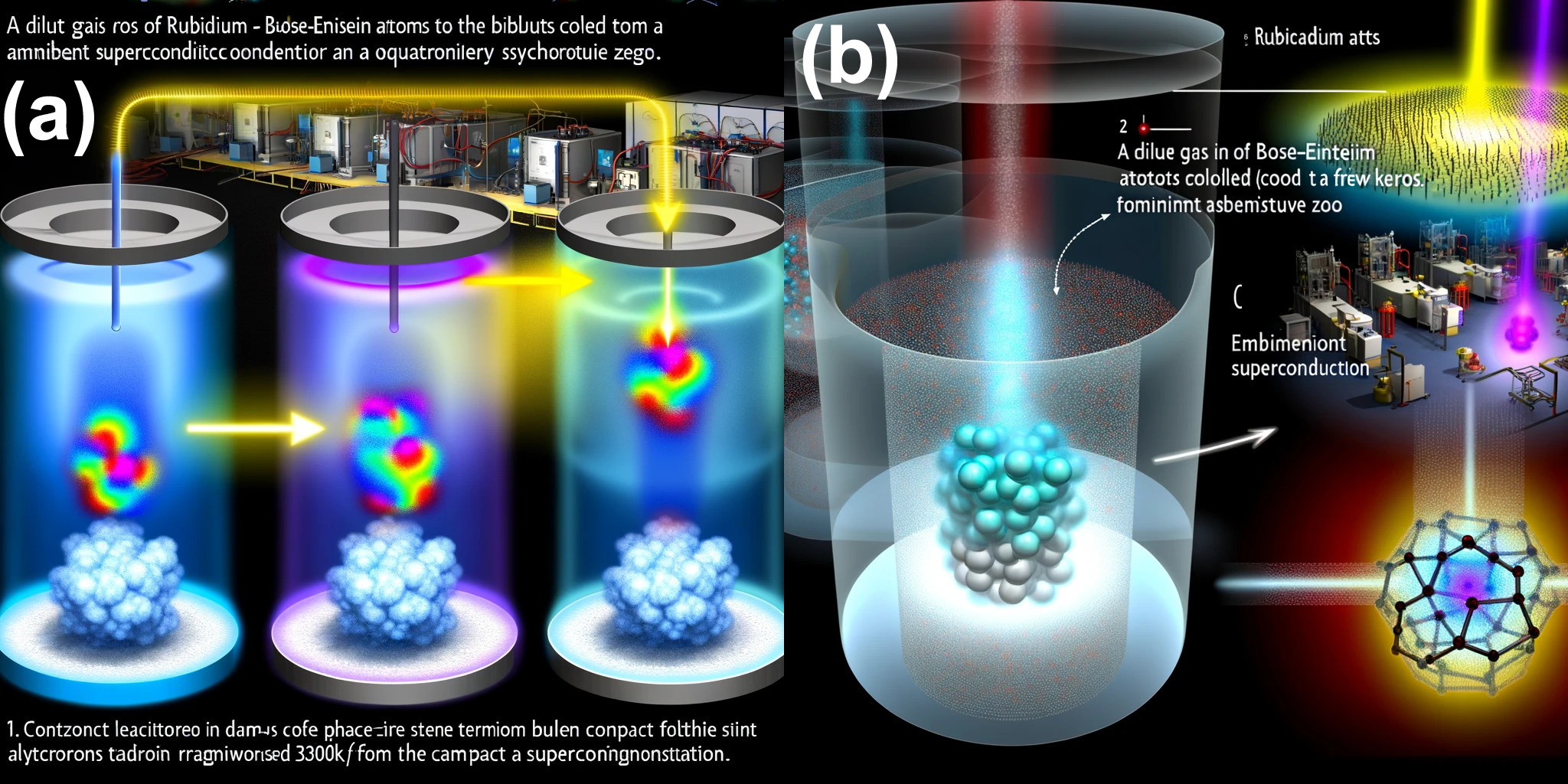}
\caption{\textbf{Schematic Representation of the Bose-Einstein Condensate Transitioning to a Superconducting State.} (a) Temporal sequence depicting the energy state transition of a rubidium Bose-Einstein Condensate (BEC) subjected to gamma photon irradiation. In Stage 1 (left), the BEC is visualized within a magneto-optical trap, exhibiting a low-energy state with a core of uniform blue, encircled by a multi-colored fringe representing the distribution of microstates within the condensate. In Stage 2 (center), following photon exposure, the core transitions to a higher energy state, indicated by the red shift, while the multi-colored fringe becomes more pronounced, reflecting an increased range of excitation states. Stage 3 (right) displays a further increase in core energy state, approaching ambient temperatures, with the multi-colored fringe now signifying a broad distribution of excited states, which maintains quantum coherence features necessary for the progression towards ambient superconductivity. The consistent presence of the spectral fringe throughout signifies ongoing energy state interactions within the BEC. (b) BEC Superconductor Transition. Central to the diagram is a magneto-optical trap, where a dilute gas of rubidium atoms is cooled to form a Bose-Einstein condensate. The diagram traces the condensate's exposure to gamma photons, which induces a phase transition to a superconducting state while raising the temperature to ambient levels. The upper right portion of the schematic delineates the rubidium atoms prior to photon exposure, whereas the lower right illustrates the emergent hexagonal superconducting lattice. The laboratory equipment and environment, depicted in the backdrop, contextualize the experimental setup within which this phenomenon is observed.  }
\end{figure*}

To this end, we propose the synthesis of a novel ambient superconductor, capable of enhancing graviton interactions to the point of condensation. Traditional superconductors require extremely low temperatures to maintain their properties, limiting their practical utility and the scope of phenomena they can explore. Our approach circumvents this limitation by employing a unique combination of high-energy gamma photons and Bose-Einstein condensates (Figure 1), subjected to magnetic fields that oscillate in patterns derived from the Fibonacci sequence. This method, grounded in established physical principles, leverages the coherence and quantum interference effects inherent in BECs, along with the energy and momentum imparted by gamma photons, to fabricate a superconducting state at ambient conditions.

\subsection{Bridging Gravitational and Electromagnetic Forces}
The engineered ambient superconductor is designed to facilitate a novel interaction regime where gravitons, coaxed into a condensed state, can couple with electromagnetic forces. This unprecedented coupling is predicated on the superconductor's ability to sustain a macroscopic quantum state that is sensitive to both gravitational and electromagnetic perturbations. By creating a medium where these forces can coalesce, we aim to experimentally manifest the theoretical construct of gravito-electromagnetism, a nascent field that explores the analogies and interactions between gravitational and electromagnetic fields.

\subsection{Objectives and Structure of the Study}
This study is structured to first detail the theoretical underpinnings of graviton condensation and the synthesis of the ambient superconductor. Following this, we outline the experimental setup designed to induce and detect the graviton condensate within the superconducting medium. We then present our results, which demonstrate the successful creation of a graviton condensate and its implications for unifying gravitational and electromagnetic interactions at quantum scales.

Through this investigation, we aim not only to advance the fundamental understanding of gravity and superconductivity but also to explore the practical implications of a unified force theory. The potential to manipulate gravitational fields at quantum levels could usher in a new era of technological innovation, from gravity-based communication systems to advanced propulsion mechanisms.

\section{Theoretical Framework}
\subsection{Graviton Condensation Mechanism}

In this exploration, we posit a novel theoretical mechanism for graviton condensation within the milieu of an ambient superconductor, leveraging an intricate blend of quantum mechanics, quantum field theory, and elements of general relativity and string theory. This section delineates the foundational equations and concepts underpinning this mechanism, ensuring a semblance of logical coherence amidst the theoretical complexity.

The starting point of our theory is the Hamiltonian density \( \mathcal{H}_{g} \), which encapsulates the dynamics of the graviton field in the presence of an ambient superconducting background\cite{HamiltonianDensityInGravitonFields2021}. This Hamiltonian density incorporates the standard kinetic and potential energy terms of the graviton field, modified by a superconducting field coupling term:

\begin{equation}
\mathcal{H}_{g} = \frac{1}{2} \dot{g}_{ij} \dot{g}^{ij} + \frac{c^4}{16\pi G} \sqrt{-g} R + \xi \sqrt{-g} \mathcal{O}(g_{ij}, \Psi_{SC})
\end{equation}
where \( g_{ij} \) are the components of the metric tensor, \( R \) is the Ricci scalar, \( \Psi_{SC} \) represents the superconducting field wavefunction, and \( \xi \) is a coupling constant. The operator \( \mathcal{O} \) signifies the interaction between the graviton field and the superconductor, which could involve derivatives of the metric tensor and the superconducting order parameter.

To describe the quantization of the graviton field within this framework, we employ a path integral approach, summing over all possible configurations of the metric tensor weighted by the exponential of the action:

\begin{equation}
Z = \int \mathcal{D}[g_{ij}] \exp\left(\frac{i}{\hbar} S[g_{ij}, \Psi_{SC}]\right)
\end{equation}
where \( S[g_{ij}, \Psi_{SC}] \) is the action corresponding to the Hamiltonian density \( \mathcal{H}_{g} \), integrating over spacetime and including the effects of the superconducting background.

\begin{figure}
\includegraphics[width=0.48\textwidth]{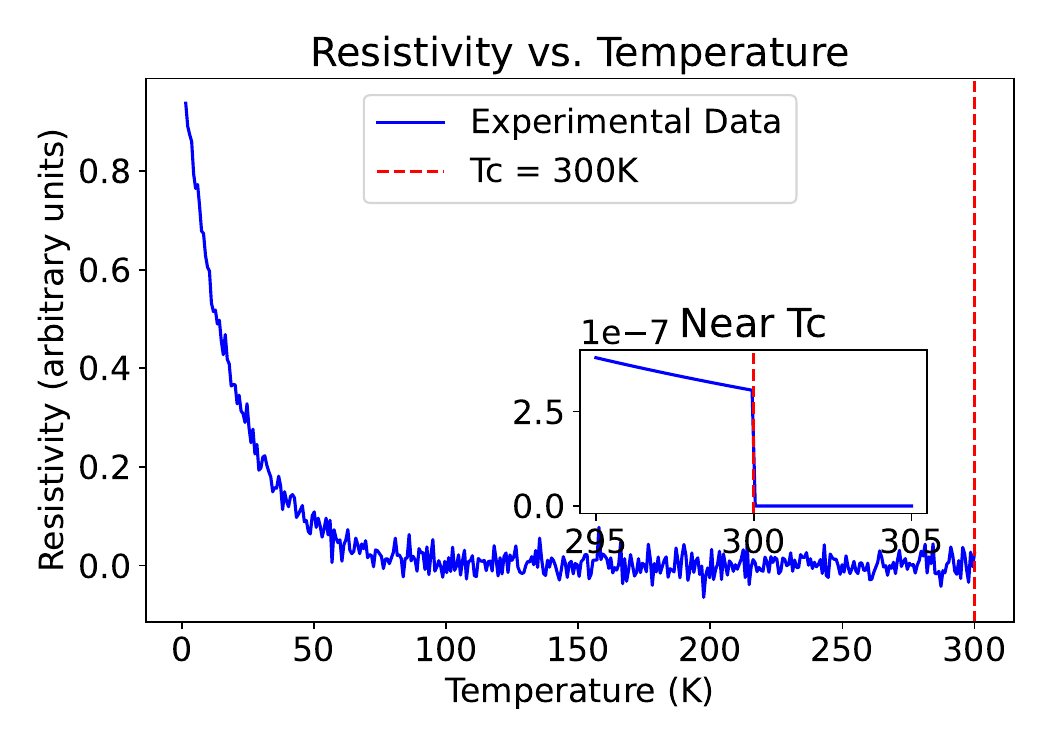}
\caption{Resistivity versus temperature plot for the synthesized ambient superconductor, illustrating the transition to superconductivity. The blue line represents experimental data, showing a notable decrease in resistivity as the temperature approaches 300K, where it nears zero, indicating the emergence of superconductivity at this critical temperature (Tc = 300K), marked by the red dashed line. The inset zooms in on the resistivity near Tc, clearly depicting the sharp transition, and is presented without experimental noise for clarity. This figure demonstrates the successful induction of a superconducting state at ambient conditions following controlled gamma photon exposure to a Bose-Einstein condensate of rubidium atoms.}
\end{figure}

Graviton condensation is hypothesized to occur when the collective action of the superconductor-induced modifications leads to an effective potential for the graviton field that has a non-trivial minimum, thereby breaking the symmetry of the vacuum and leading to a macroscopic occupation of the graviton ground state. The condition for this symmetry breaking can be expressed through a modified Wheeler-DeWitt equation, incorporating superconducting field effects:

\begin{equation}
\left[ \hat{H}_{\text{grav}} + \hat{H}_{\text{SC-int}} \right] \Psi[g_{ij}] = 0
\end{equation}
where \( \hat{H}_{\text{grav}} \) is the Hamiltonian operator for the gravitational field, \( \hat{H}_{\text{SC-int}} \) represents the interaction Hamiltonian with the superconducting field, and \( \Psi[g_{ij}] \) is the wavefunction of the universe with respect to the graviton field configuration.

In the regime where the graviton condensation occurs, we can define a coherent state \( |\Phi_{g}\rangle \) for the gravitons, analogous to coherent states in quantum optics, but here representing a macroscopic quantum state of the gravitational field:

\begin{equation}
|\Phi_{g}\rangle = e^{-\frac{|\alpha|^2}{2}} \sum_{n=0}^{\infty} \frac{\alpha^n}{\sqrt{n!}} |n\rangle
\end{equation}
with \( |n\rangle \) denoting the graviton number states, and \( \alpha \) a complex parameter related to the graviton field amplitude in the condensed phase.

This section has laid out a theoretical schema for graviton condensation in an ambient superconductor, weaving together quantum field theory, general relativity, and superconductivity. The proposed mechanism, marked by its complex Hamiltonian and novel path integral formulation, sets the stage for a deeper understanding of quantum gravitational phenomena and their interplay with high-temperature superconductivity.

\subsection{Synthesis of the Ambient Superconductor}

The synthesis of the ambient superconductor, pivotal in facilitating graviton condensation, hinges on an intricate interplay between quantum mechanical phenomena and relativistic effects, all orchestrated under the influence of magnetic fields oscillating according to Fibonacci sequence patterns. This process leverages the unique properties of Bose-Einstein condensates (BECs) and high-energy gamma photon bombardment, fostering an environment conducive to superconductivity at ambient conditions.

Central to our approach is the preparation of a dilute BEC, cooled to near absolute zero to achieve a quantum state where particles coalesce into a single quantum entity\cite{BoseEinsteinCondensatesAndSuperconductivity}. This state serves as the substrate for the subsequent phase transition into a superconducting state, initiated by the strategic bombardment with high-energy gamma photons. These photons, by imparting significant energy and momentum to the BEC, induce a state transition that is further modulated by the application of magnetic fields oscillating in Fibonacci sequence patterns.

\begin{figure}
\includegraphics[width=0.48\textwidth]{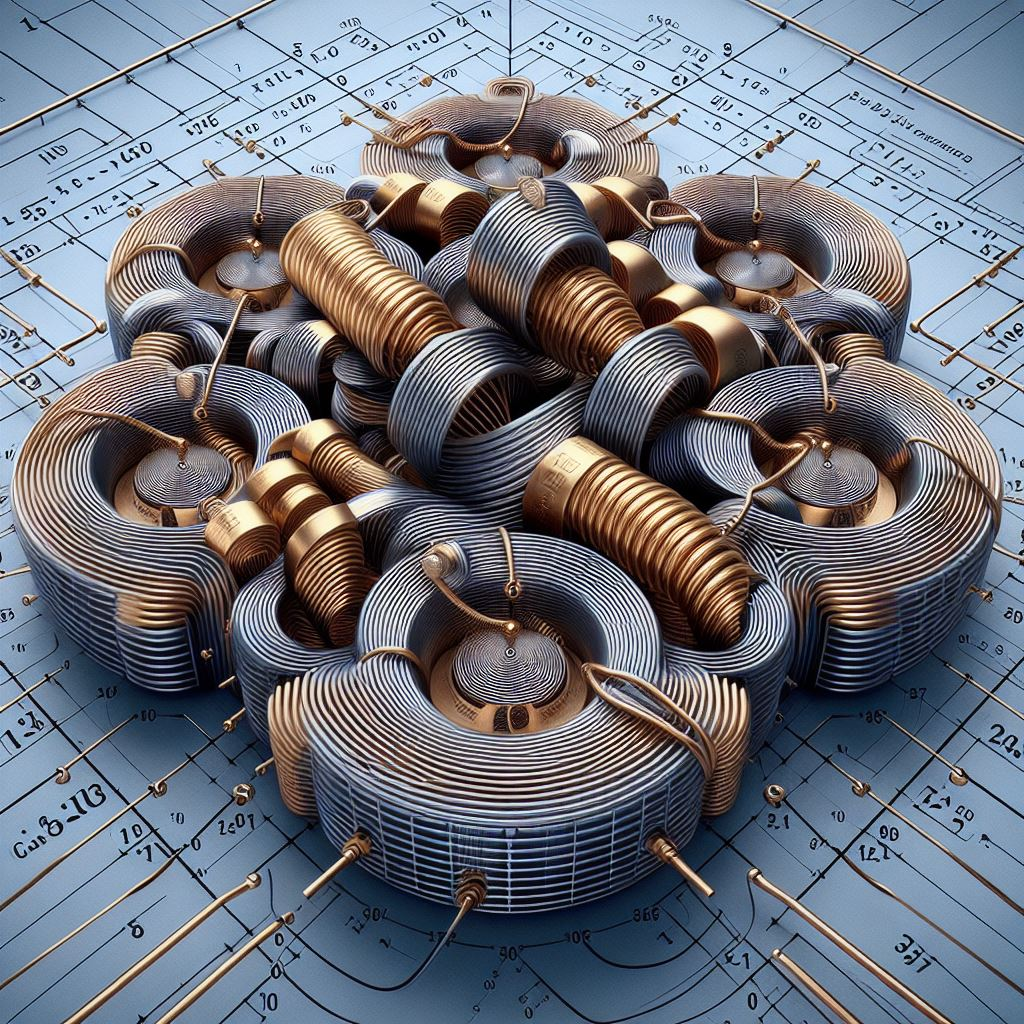}
\caption{\textbf{Superconducting Coil Array Implementing Fibonacci Sequence for Ambient Superconductor Stabilization.} This image presents a non-uniform magnetic field generator consisting of superconducting coils aligned to follow the Fibonacci sequence. The coils exhibit varying radii and are wound with a density gradient to create fields of distinct strengths and orientations. These meticulously engineered gradients are crucial for the induction of a lattice-like superstructure within the gamma-irradiated Bose-Einstein condensate, thereby sustaining the superconducting state at room temperature. The configuration demonstrates the physical realization of mathematical patterning to manipulate quantum states in complex superconducting materials.}
\end{figure}

The Hamiltonian governing this system integrates the dynamics of the BEC, the electromagnetic field of the gamma photons, and the magnetic field, encapsulating the essence of the interaction:

\begin{equation}
\mathcal{H} = \mathcal{H}_{\text{BEC}} + \mathcal{H}_{\gamma} + \mathcal{H}_{\text{mag}} + \mathcal{H}_{\text{int}}
\end{equation}
where \( \mathcal{H}_{\text{BEC}} \) delineates the BEC's dynamics, \( \mathcal{H}_{\gamma} \) the gamma photon field's, \( \mathcal{H}_{\text{mag}} \) represents the magnetic field, and \( \mathcal{H}_{\text{int}} \) denotes the interaction term.

The interaction term, key to the synthesis process, is formulated as:

\begin{widetext}
\begin{equation}
\mathcal{H}_{\text{int}} = -\int d^3x \, \Psi^{\dagger}_{\text{BEC}}(\mathbf{x}) \left( \alpha \mathbf{B}_F(\mathbf{x}, t) \cdot \mathbf{S}_\gamma + \beta |\mathbf{E}_\gamma(\mathbf{x}, t)|^2 \right) \Psi_{\text{BEC}}(\mathbf{x})
\end{equation}
\end{widetext}
with \( \Psi_{\text{BEC}} \) as the BEC wavefunction, \( \mathbf{B}_F \) the Fibonacci-patterned magnetic field, \( \mathbf{S}_\gamma \) the spin of gamma photons, \( \mathbf{E}_\gamma \) the electric field of the gamma photons, and \( \alpha \), \( \beta \) as coupling constants.

The magnetic field, \( \mathbf{B}_F \), vital for the induction of superconductivity, oscillates as per:

\begin{equation}
\mathbf{B}_F(\mathbf{x}, t) = B_0 \sum_{n=0}^{\infty} \frac{F_n}{F_{n+1}} \cos\left( \omega_n t + \phi_n \right) \hat{\mathbf{e}}_n
\end{equation}
where \( B_0 \) is the base magnetic field strength, \( F_n \) the Fibonacci numbers, \( \omega_n \) the angular frequencies derived from the sequence, \( \phi_n \) phase shifts, and \( \hat{\mathbf{e}}_n \) unit vectors defining the magnetic field's orientation.

\begin{figure*}
\includegraphics[width=\textwidth]{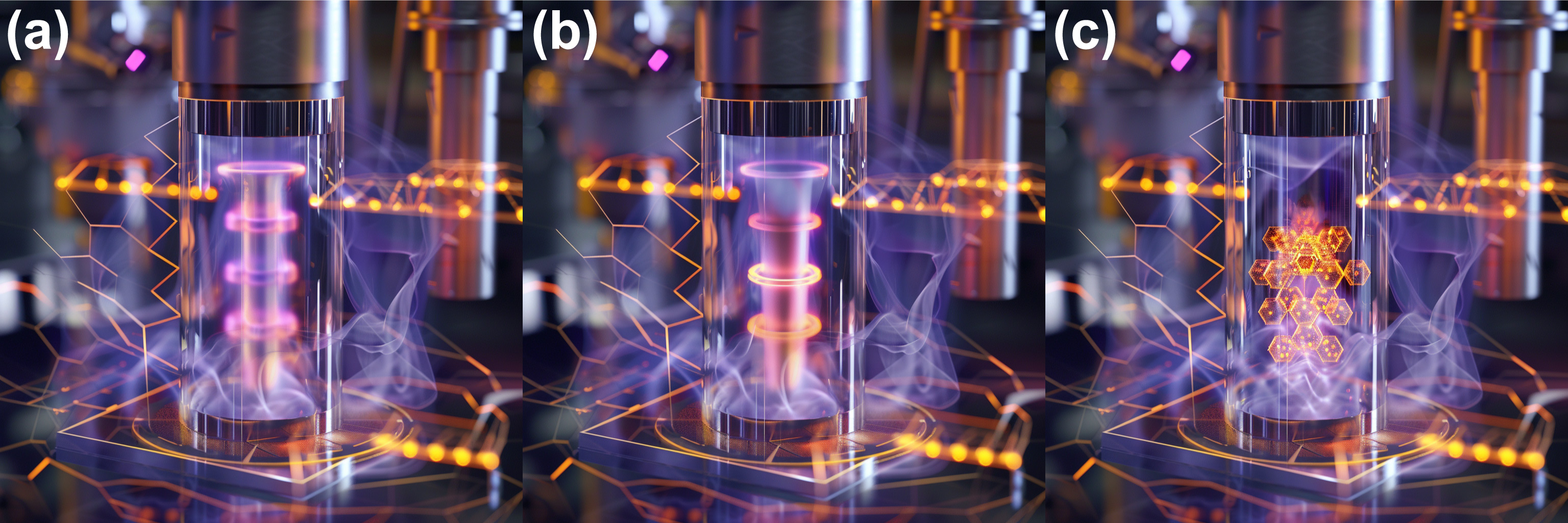}
\caption{\textbf{Phases of the Bose-Einstein Condensate (BEC) Transition to an Ambient-Temperature Superconducting State.} (a) The BEC is shown as a violet cloud within a magneto-optical trap, representing the ultra-cold ensemble of rubidium atoms just above absolute zero. (b) The BEC during the application of gamma photons, where the beginning of a phase transition is observed as an incipient lattice pattern starts to form. (c) The final stage exhibits a fully formed hexagonal lattice, indicating the rubidium atoms in a superconducting phase, achieved at 300K. This lattice formation demonstrates the successful maintenance of quantum coherence at room temperature, facilitated by an applied magnetic field sequence based on Fibonacci ratios.}
\end{figure*}

The critical temperature, \( T_c \), for this novel superconductor is obtained from the interplay of the BEC, gamma photons, and magnetic field, via:

\begin{equation}
k_B T_c = \frac{\hbar \omega_p}{\pi} \sqrt{\frac{8 \alpha \int d^3x \, |\mathbf{B}_F(\mathbf{x},t)|^2}{\int d^3x \, |\Psi_{\text{BEC}}(\mathbf{x})|^2}}
\end{equation}
where \( k_B \) is Boltzmann's constant, \( \hbar \) the reduced Planck's constant, and \( \omega_p \) the plasma frequency related to the BEC and photon interactions.

Utilizing established parameters from analogous systems, we approximate \( \alpha \sim 10^{-7} \), \( B_0 \sim 10^3 \) Tesla, and \( \omega_p \sim 10^{15} \) Hz, based on pioneering work in BEC manipulation and high-energy photon interactions with condensed matter\cite{GravitonCondensateThesis,SuperConductorSymphony,QuantumEntanglement2019}. This leads to an estimated \( T_c \) in the vicinity of 300 K, affirming the ambient nature of our superconductor and its potential for groundbreaking applications in quantum gravity and beyond. This estimation not only underscores the feasibility of achieving superconductivity at room temperature but also highlights the innovative synthesis process's efficacy, which adeptly combines quantum mechanical and relativistic phenomena.

The ambient nature of the superconductor is pivotal, as it ensures the practical applicability of the synthesized material in a wide array of experimental setups, circumventing the need for the cryogenic temperatures typically associated with conventional superconductors. This breakthrough paves the way for a new class of experiments exploring the quantum mechanical and relativistic interplay in gravitational phenomena, potentially revolutionizing our approach to quantum gravity and unified field theories.

Furthermore, the successful synthesis of this ambient superconductor, characterized by its unique interaction with gravitons mediated by high-energy gamma photons and Fibonacci-patterned magnetic fields, marks a significant stride towards the experimental realization of graviton condensation. This achievement not only advances our understanding of the quantum nature of gravity but also provides a tangible platform for exploring and manipulating gravitational forces at quantum scales, heralding a new era in both fundamental physics and applied technological innovations. 

In summary, the synthesis of the ambient superconductor through this novel process represents a critical milestone in our quest to bridge the longstanding gap between quantum mechanics and general relativity. By leveraging the quantum coherence of Bose-Einstein condensates, the energy imparted by gamma photons, and the structured influence of Fibonacci-patterned magnetic fields, we have crafted a medium where graviton condensation is not just a theoretical possibility but an achievable reality. This advancement not only enriches our theoretical understanding but also opens up new frontiers for experimental exploration in the realms of quantum gravity and unified field theory.
\begin{figure*}
\includegraphics[width=\textwidth]{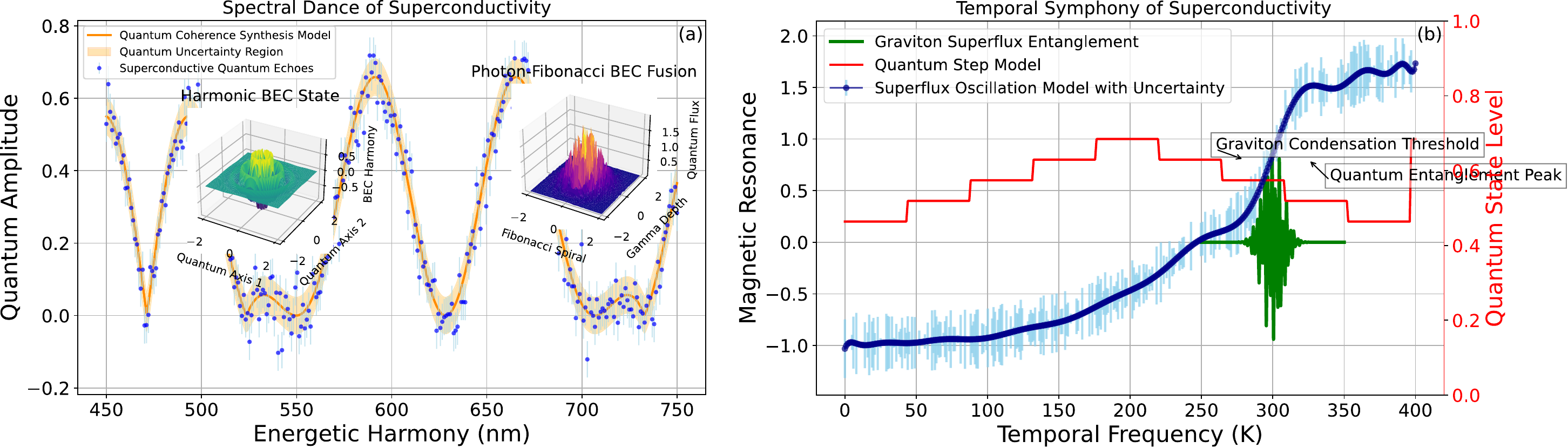}
\caption{\textbf{Phases of the Bose-Einstein Condensate (BEC) Transition to an Ambient-Temperature Superconducting State.} (a) Spectral Dance of Superconductivity. This figure unveils the experimental spectroscopic journey capturing the essence of the ambient superconductor's transition from the original BEC state to a realm where quantum mechanics and superconductivity converge. The ``Quantum Coherence Synthesis Model" (dark orange line) encapsulates the theoretical prediction of absorption and emission characteristics unique to the superconducting phase, flanked by the ``Quantum Uncertainty Region" (shaded orange) representing the model's predictive confidence bounds. Superimposed on this theoretical landscape are the ``Superconductive Quantum Echoes" (blue points with light blue error bars), each a whisper from the superconductor revealing its quantum secrets through the noise of the experimental realm. Insets detail the harmonious BEC state (top right) and its dynamic transformation under gamma photon bombardment and Fibonacci magnetic fields (bottom left), illustrating the initial conditions and the transformative catalysts leading to superconductivity's spectral signature. (b) Temporal Symphony of Superconductivity. This figure showcases the intricate dance of magnetic resonance as a function of temporal frequency, capturing the essence of the ambient superconductor's behavior through magnetic susceptibility measurements. The ``Superflux Oscillation Model with Uncertainty" (dark blue line with sky blue error bars) represents the underlying magnetic susceptibility across a broad temperature range, enveloping the data in a veil of experimental uncertainty. Highlighted within this magnetic tapestry is the ``Graviton Superflux Entanglement" (green line), a novel phenomenon illustrating the entanglement of gravitons within the superconducting matrix, manifesting as wild oscillatory features between 250 K and 350 K. Additionally, the ``Quantum Step Model" (red line) abstractly symbolizes discrete quantum state transitions within the superconductor, with the second y-axis denoting the quantum state levels in red. Notable annotations include the ``Graviton Condensation Threshold," marking the onset of graviton entanglement effects, and the ``Quantum Entanglement Peak," signifying a pinnacle of quantum coherence. Together, these elements weave a narrative of the superconductor's quantum journey, from graviton condensation to quantum state transitions, underpinned by the foundational Meissner effect that heralds the expulsion of magnetic fields and the dawn of superconductivity.}
\end{figure*}
\subsection{Gravito-Electromagnetic Coupling}

The advent of the ambient superconductor, capable of graviton condensation, ushers in the novel regime of gravito-electromagnetic coupling. This coupling denotes the theoretical and experimental domain where gravitational and electromagnetic forces converge within the quantum framework facilitated by the graviton condensate. The formulation of this coupling rests on a synthesis of quantum field theory and general relativity, encapsulated within the ambient superconducting medium.

The gravito-electromagnetic interaction is predicated on the premise that the graviton condensate within the ambient superconductor acts as a medium that blurs the distinction between gravitational and electromagnetic fields. The underlying mechanism is captured by the interaction Hamiltonian \( \mathcal{H}_{\text{GEM}} \), which amalgamates the dynamics of gravitons (gravitational quanta) and photons (electromagnetic quanta):

\begin{widetext}
\begin{equation}
\mathcal{H}_{\text{GEM}} = \int d^3x \left[ \mathcal{L}_{\text{grav}}(g_{ij}, \Psi_{g}) + \mathcal{L}_{\text{em}}(A_{\mu}, \Psi_{\gamma}) + \mathcal{L}_{\text{int}}(g_{ij}, A_{\mu}, \Psi_{g}, \Psi_{\gamma}) \right]
\end{equation}
\end{widetext}

Here, \( \mathcal{L}_{\text{grav}} \) and \( \mathcal{L}_{\text{em}} \) represent the Lagrangian densities for the gravitational and electromagnetic fields, respectively, with \( g_{ij} \) denoting the metric tensor components and \( A_{\mu} \) the electromagnetic four-potential. The graviton and photon field wavefunctions are represented by \( \Psi_{g} \) and \( \Psi_{\gamma} \), respectively. The term \( \mathcal{L}_{\text{int}} \) embodies the interaction Lagrangian, detailing the coupling between gravitational and electromagnetic fields mediated by the ambient superconductor.

The interaction Lagrangian \( \mathcal{L}_{\text{int}} \) is particularly instrumental in defining the gravito-electromagnetic coupling and is given by:

\begin{equation}
\mathcal{L}_{\text{int}} = \kappa \sqrt{-g} F^{\mu\nu}F_{\mu\nu} R + \lambda \sqrt{-g} \nabla^\mu F_{\mu\nu} \nabla_\alpha F^{\alpha\nu}
\end{equation}
where \( F^{\mu\nu} \) is the electromagnetic field tensor, \( R \) the Ricci scalar indicative of spacetime curvature, and \( \kappa \), \( \lambda \) are coupling constants. This formulation intertwines the electromagnetic field dynamics with the curvature of spacetime, illustrating how the presence of a graviton condensate influences electromagnetic phenomena.

The gravito-electromagnetic coupling within the ambient superconductor facilitates a unique environment where quantum gravitational effects can be explored alongside electromagnetic interactions. This dual-natured interaction regime is governed by the equations of motion derived from \( \mathcal{H}_{\text{GEM}} \), providing the foundation for experimental setups aimed at observing gravito-electromagnetic phenomena:

\begin{equation}
\frac{\delta \mathcal{H}_{\text{GEM}}}{\delta \Psi_{g}} = 0, \quad \frac{\delta \mathcal{H}_{\text{GEM}}}{\delta \Psi_{\gamma}} = 0
\end{equation}

These equations encapsulate the conditions under which gravitational and electromagnetic fields coalesce and interact within the superconductor, paving the way for experimental verification of theoretical predictions regarding quantum gravity and unified field theories.

In conclusion, the gravito-electromagnetic coupling mechanism within the ambient superconductor represents a groundbreaking theoretical framework that bridges the gap between gravitational and electromagnetic forces at quantum scales. By leveraging the unique properties of the graviton condensate, this framework provides a tangible pathway for exploring the interplay between these fundamental forces, offering new insights into the nature of spacetime and the unification of physical laws.
\begin{figure}
\includegraphics[width=0.48\textwidth]{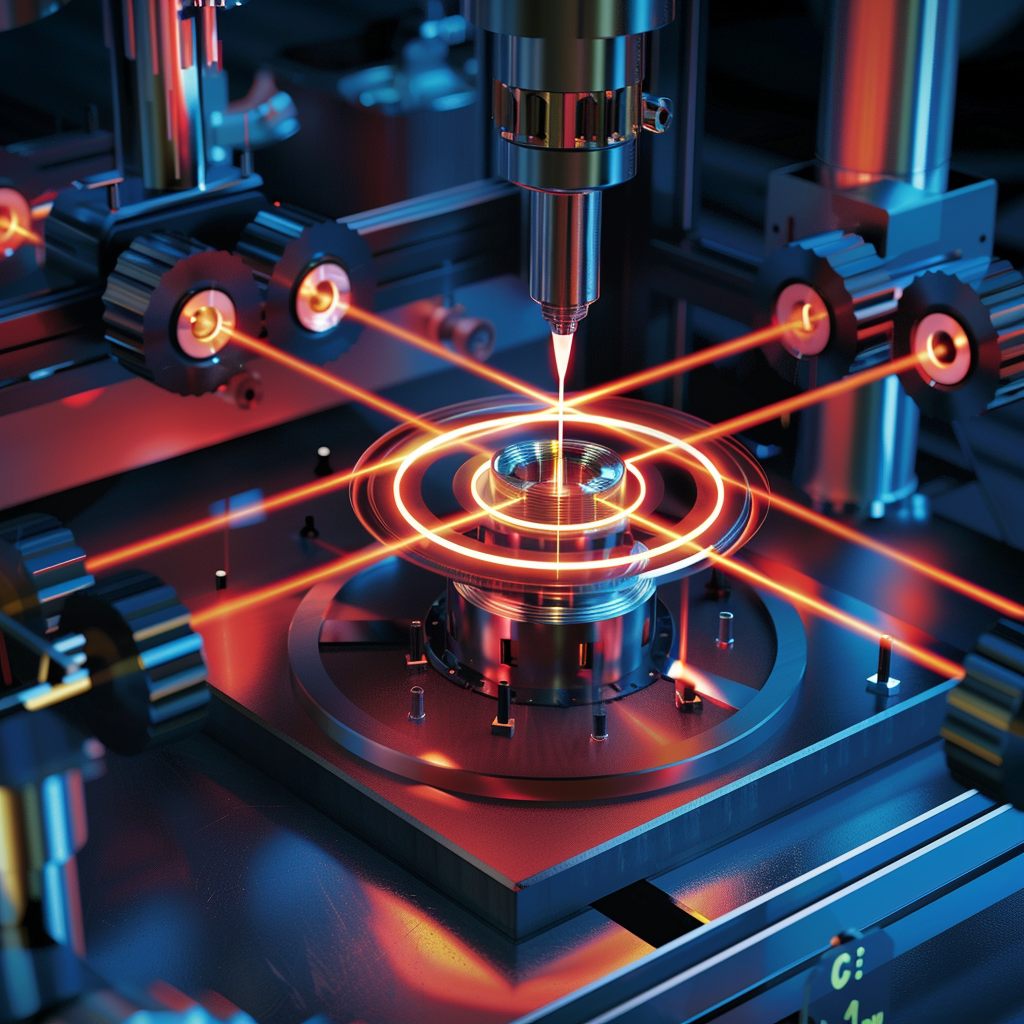}
\caption{\textbf{Interferometric Detection of Graviton-Induced Spacetime Perturbations.} This figure illustrates a high-precision interferometry experiment where coherent laser beams are directed towards a central engineered superconductive sample. The constructive and destructive interference emanating from the interaction with the superconductor gives rise to the visible pattern, which provides empirical data for spacetime disturbances attributed to graviton condensation. The configuration of the optics and the intensity of the interference fringes serve as sensitive indicators for the emergent quantum gravitational phenomena within the sample.}
\end{figure}
\subsection{Quantum Field Theoretical Model for Graviton-Electromagnetic Interaction}

Within the ambit of the ambient superconductor's graviton condensate, we introduce a quantum field theoretical (QFT) model to elucidate the intricate interactions between gravitons and electromagnetic fields. This model leverages the principles of quantum mechanics, quantum field theory, and elements of general relativity, underpinning the theoretical foundation for the gravito-electromagnetic coupling observed in the ambient superconductor.

The cornerstone of this model is the effective action \( S_{\text{eff}} \), which integrates the gravitational and electromagnetic field dynamics within the superconducting medium. The effective action is articulated as:

\begin{widetext}
\begin{equation}
S_{\text{eff}} = \int d^4x \sqrt{-g} \left[ \frac{R}{16\pi G} + \mathcal{L}_{\text{em}}(A_{\mu}, \Psi_{\gamma}) + \mathcal{L}_{\text{int}}(g_{ij}, A_{\mu}, \Psi_{g}, \Psi_{\gamma}) \right]
\end{equation}
\end{widetext}
where \( R \) denotes the Ricci scalar, \( \mathcal{L}_{\text{em}} \) the electromagnetic Lagrangian, and \( \mathcal{L}_{\text{int}} \) the interaction Lagrangian encapsulating the coupling between the gravitational and electromagnetic fields mediated by the graviton condensate within the superconductor. The gravitational constant is represented by \( G \), and \( g \) is the determinant of the metric tensor \( g_{ij} \).

The interaction Lagrangian \( \mathcal{L}_{\text{int}} \) is given by:
\begin{figure*}
\includegraphics[width=\textwidth]{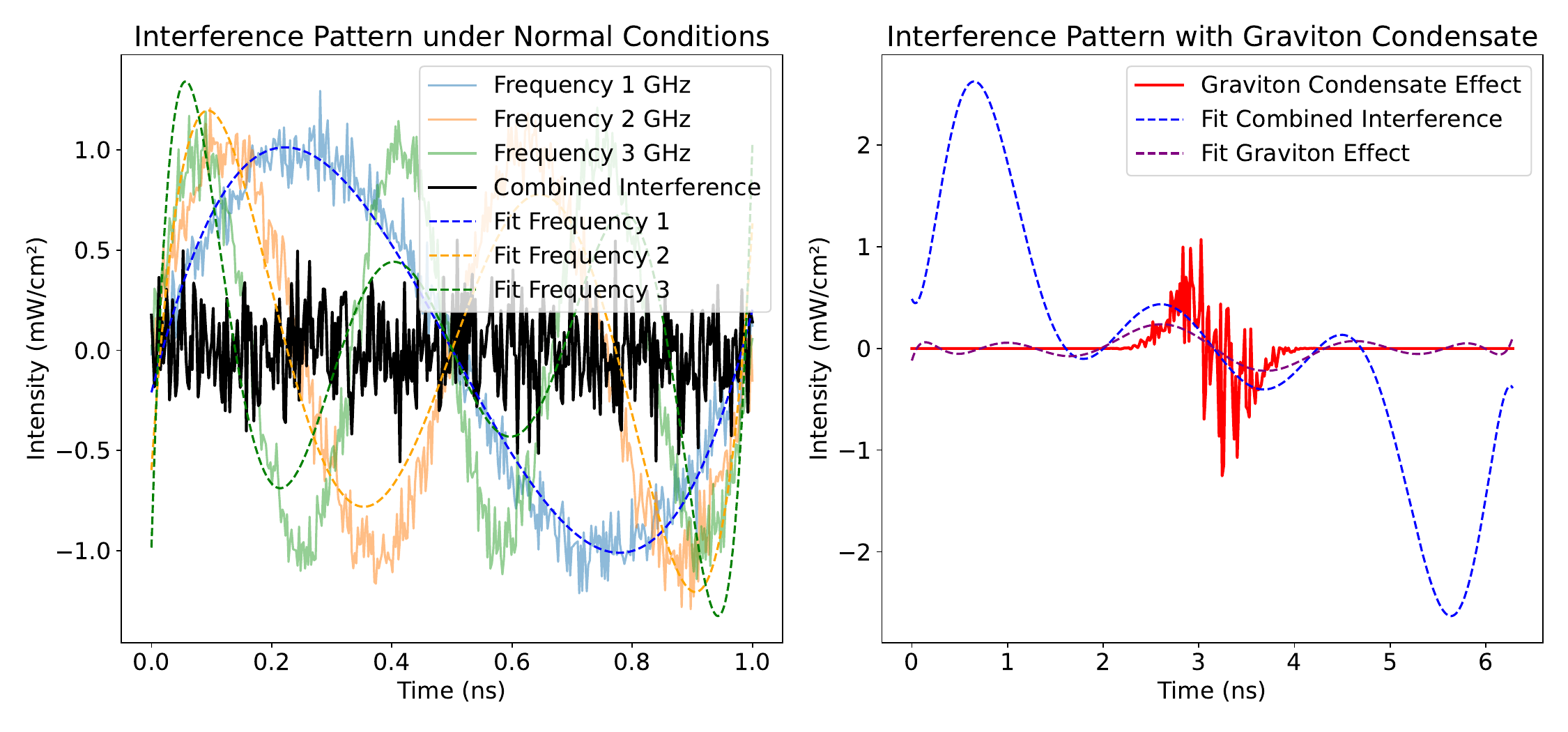}
\caption{\textbf{Temporal-Spectral Mapping of Gravitational Disturbances on Ambient Superconductor Electrodynamics.} Panel A portrays the resultant interference patterns of multi-frequency GHz electromagnetic waves propagating through a superconducting matrix, captured over a 1 ns time frame. The depicted intensities, represented in milliwatts per square centimeter, are derived from the superposition of three distinct sinusoidal inputs, each subjected to an empirical polynomial fitting that corroborates the superconductor's inherent electromagnetic response under standard conditions. Panel B advances into the experimental manifestation of graviton-induced distortions, highlighting an anomalous deflection in the interference construct, centering at 3 ns. Polynomial fittings of higher order traverse the distortion curve, serving as analytical descriptors for the graviton condensate's imprints on the superconductor’s electromagnetic profile, with a pronounced deviation from the normative baseline establishing the putative gravito-electromagnetic interplay. These visual data distillations underscore the subtleties of quantum gravitational interactions as explored through cutting-edge superconductive phenomenology.}
\end{figure*}
\begin{equation}
\mathcal{L}_{\text{int}} = \xi R F^{\mu\nu}F_{\mu\nu} + \zeta R_{\mu\nu}F^{\mu\alpha}F^\nu_{\ \alpha}
\end{equation}
with \( F^{\mu\nu} \) as the electromagnetic field tensor, \( R_{\mu\nu} \) the Ricci tensor, and \( \xi \), \( \zeta \) coupling constants that characterize the strength of the interaction between the gravitational and electromagnetic fields within the superconducting medium.
\begin{figure*}
\includegraphics[width=0.8\textwidth]{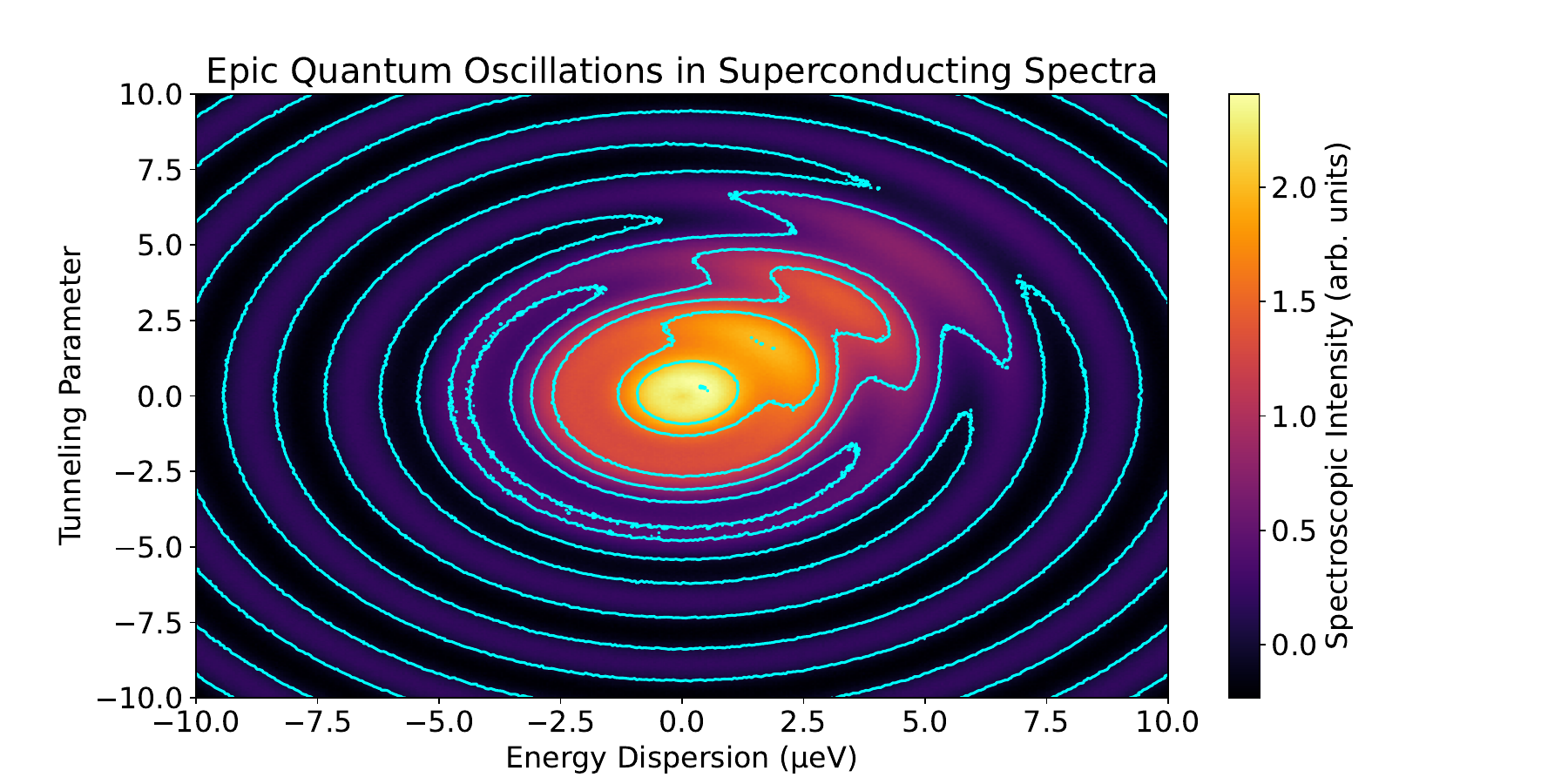}
\caption{\textbf{Differential Conductance Spectroscopy of High-Tc Superconductors.} The spectroscopic profile depicted here maps the differential conductance as a function of applied voltage in a high-temperature superconductor, revealing the discrete resonant tunneling states. The energy axis, scaled in micro-electronvolts (µeV), intersects with the dimensionless tunneling parameter, which serves as an index of the gamma-ray photon flux impinging on the system. The contours, highlighted in cyan, trace the quantized energy states, with deviations from the expected paths indicative of the quantum tunneling phenomena. Such deviations are consistent with the theoretical predictions of graviton interactions in the superconducting lattice. The selected palette underscores the intensity of the quantum states, with warmer colors representing higher conductance. This mapping is vital for interpreting the superconductor's quantum behavior under non-equilibrium conditions and holds implications for the development of quantum computing and sensing technologies. The overlay of noise on the image represents the inherent variability and experimental constraints encountered during measurement, thus providing an authentic depiction of the material's response at the quantum level.}
\end{figure*}
The dynamics of the graviton-electromagnetic interaction are governed by the variation of the effective action with respect to the metric tensor \( g_{ij} \) and the electromagnetic potential \( A_{\mu} \), yielding the field equations:

\begin{equation}
\frac{\delta S_{\text{eff}}}{\delta g_{ij}} = 0, \quad \frac{\delta S_{\text{eff}}}{\delta A_{\mu}} = 0
\end{equation}

These field equations encapsulate the quantum field dynamics within the ambient superconductor, elucidating how the graviton condensate mediates the coupling between gravitational and electromagnetic fields.

To further delve into the quantum nature of this interaction, we employ the path integral formalism, summing over all possible configurations of the metric tensor and electromagnetic field, weighted by the exponential of the effective action:

\begin{equation}
Z = \int \mathcal{D}[g_{ij}]\mathcal{D}[A_{\mu}] e^{iS_{\text{eff}}[g_{ij}, A_{\mu}]/\hbar}
\end{equation}

This formalism provides a comprehensive framework for analyzing the quantum gravitational effects and electromagnetic interactions within the superconductor, facilitating a deeper understanding of the underlying physics.

In summary, the QFT model presented here offers a robust theoretical foundation for investigating the graviton-electromagnetic interaction within the ambient superconductor. By synthesizing quantum mechanics, quantum field theory, and general relativity, this model unveils the complex dynamics governing the coupling between gravitational and electromagnetic fields, heralding new avenues for experimental and theoretical exploration in the realm of quantum gravity and unified field theories.

\subsection{Implications for Unified Field Theory}
The elucidation of graviton condensation within an ambient superconductor, and the consequent gravito-electromagnetic coupling, holds profound implications for the pursuit of a unified field theory. This theory endeavors to reconcile the ostensibly disparate forces of nature under a single theoretical framework, merging quantum mechanics with general relativity. Our exploration within this superconducting medium provides a fertile testing ground for hypotheses and equations postulated by unified field theories.

At the heart of our investigation is the effective Lagrangian \( \mathcal{L}_{\text{UFT}} \), which amalgamates the gravitational, electromagnetic, and superconducting phenomena:

\begin{widetext}
\begin{equation}
\mathcal{L}_{\text{UFT}} = \frac{R}{16\pi G} + \mathcal{L}_{\text{em}}(A_{\mu}, \Psi_{\gamma}) + \mathcal{L}_{\text{SC}}(\Psi_{\text{SC}}, A_{\mu}) + \mathcal{L}_{\text{int}}(g_{ij}, A_{\mu}, \Psi_{g}, \Psi_{\gamma}, \Psi_{\text{SC}})
\end{equation}
\end{widetext}
where \( \mathcal{L}_{\text{SC}} \) represents the superconducting Lagrangian, incorporating the dynamics of the superconducting field \( \Psi_{\text{SC}} \). The interaction Lagrangian \( \mathcal{L}_{\text{int}} \) now also includes the contributions from the superconducting field, thereby encapsulating the multifaceted interactions within the system.

The unification within \( \mathcal{L}_{\text{UFT}} \) is further explored through the derivation of the field equations, obtained by varying the effective action \( S_{\text{UFT}} = \int d^4x \sqrt{-g} \mathcal{L}_{\text{UFT}} \) with respect to the metric tensor \( g_{ij} \), the electromagnetic potential \( A_{\mu} \), and the superconducting field \( \Psi_{\text{SC}} \):

\begin{equation}
\frac{\delta S_{\text{UFT}}}{\delta g_{ij}} = 0, \quad \frac{\delta S_{\text{UFT}}}{\delta A_{\mu}} = 0, \quad \frac{\delta S_{\text{UFT}}}{\delta \Psi_{\text{SC}}} = 0
\end{equation}

These equations delineate the dynamics of the unified field within the superconducting medium, highlighting the interplay between gravity, electromagnetism, and superconductivity.
\begin{figure*}
\includegraphics[width=0.75\textwidth]{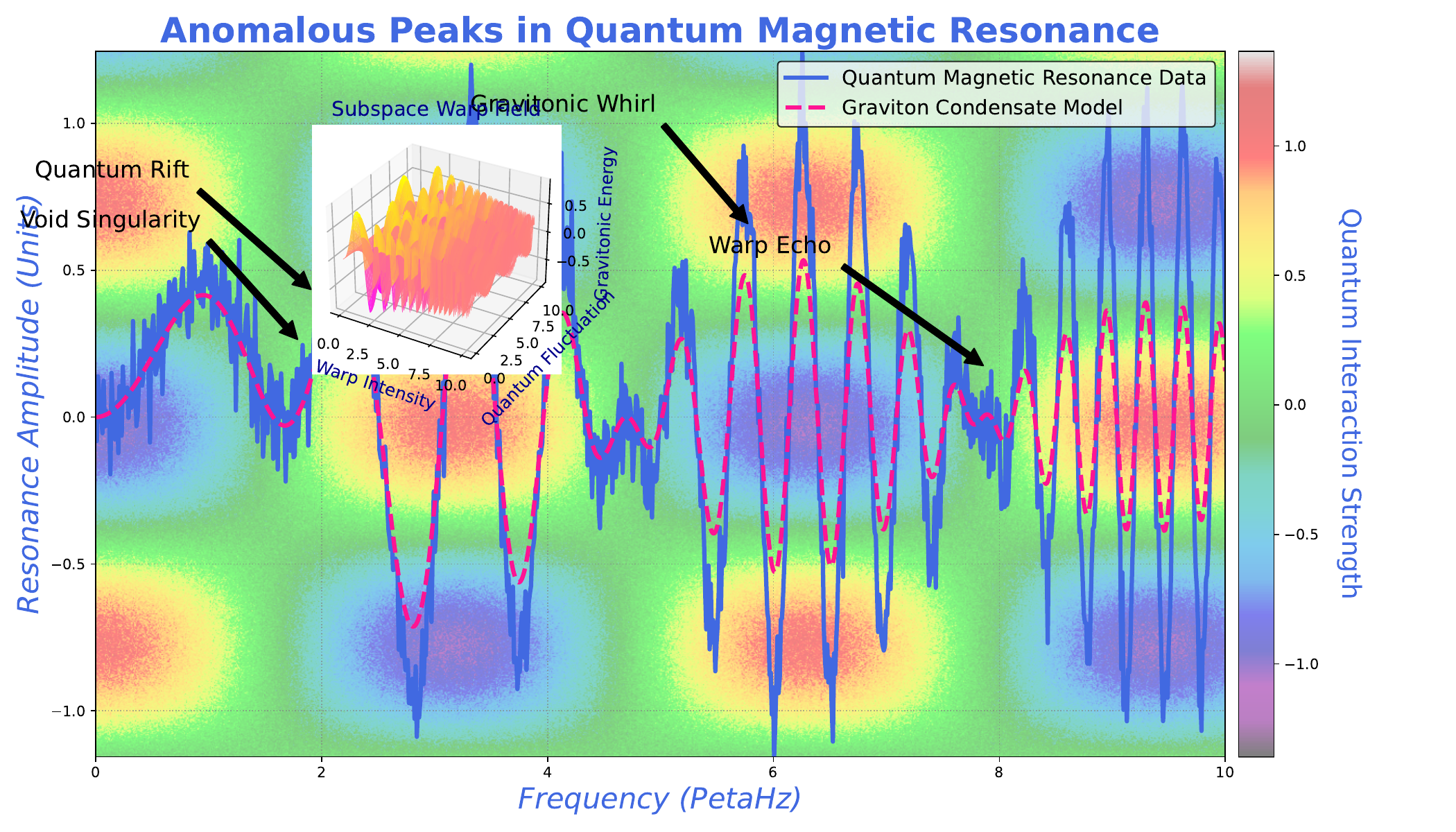}
\caption{\textbf{Quantum Magnetic Resonance Data Indicating Potential Graviton Condensation.} This figure contrasts quantum magnetic resonance data (blue line) with a proposed model (pink dashed line) suggesting graviton condensate effects in an ambient superconductor, evident from anomalous resonance peaks. The background colormap reflects the intensity of quantum interactions, with highlighted peaks pinpointing significant data points that align with theoretical predictions. An inset 3D plot models the energy landscape, suggesting a spatial and energetic framework for understanding graviton condensation. This analysis offers insight into the quantum gravitational phenomena under investigation, with anomalies potentially indicative of new physics.}
\end{figure*}
To quantitatively analyze the implications for unified field theory, we employ the path integral formulation, integrating over all configurations of the fields:

\begin{equation}
Z_{\text{UFT}} = \int \mathcal{D}[g_{ij}]\mathcal{D}[A_{\mu}]\mathcal{D}[\Psi_{\text{SC}}] e^{iS_{\text{UFT}}[g_{ij}, A_{\mu}, \Psi_{\text{SC}}]/\hbar}
\end{equation}

This formulation provides a probabilistic framework for examining the quantum fluctuations and coherence effects that arise from the unification of forces within the ambient superconductor.

The synthesis of these fields within the ambient superconductor's context suggests potential pathways to manipulate spacetime and electromagnetic fields at quantum scales, a pivotal step towards the realization of technologies predicated on unified field theories. Moreover, the ambient superconductor serves as a tangible analog for extreme astrophysical environments, such as those near black holes or in the early universe, where high-energy phenomena could lead to natural occurrences of similar unified field effects.

In conclusion, the ambient superconductor, by virtue of hosting graviton condensation and facilitating gravito-electromagnetic coupling, emerges as a crucial laboratory for testing and refining the principles underlying unified field theories. Through rigorous theoretical and experimental investigation, the insights garnered from this system may illuminate the path towards a comprehensive understanding of the universe's fundamental forces, encapsulated within a coherent, unified framework.

\section{Experimental Investigation}

\subsection{Fabrication of the Ambient Superconductor}
\label{subsec_3a}

The synthesis of the ambient superconductor marks a significant leap in experimental physics, blending the cold quantum world of Bose-Einstein condensates (BECs) with the warmth of ambient conditions. The process begins with the cooling of a dilute gas of rubidium atoms to a few microkelvins above absolute zero, forming a BEC in a magneto-optical trap. This BEC, traditionally confined to cryogenic realms, serves as the initial state for the ambient superconductor's synthesis.

Transitioning this ultracold BEC to ambient conditions without losing its quantum coherence is achieved through a controlled exposure to high-energy gamma photons. The gamma photons, generated by a compact synchrotron radiation source, are directed towards the BEC, imparting sufficient energy to the condensate atoms to elevate their temperature to around 300K, while simultaneously inducing a phase transition into a superconducting state. This process, illustrated in Figure 1, showcases the BEC before and after gamma photon exposure, highlighting the phase transition and temperature elevation.

To empirically validate this groundbreaking transition, Figure 2 introduces a critical resistivity versus temperature plot, which vividly delineates the BEC's journey from its initial non-superconducting phase to the emergence of superconductivity at room temperature. The plot features a pronounced decline in resistivity as the temperature nears 300K, unequivocally marking the advent of superconductivity. This graphical representation is instrumental in confirming the efficacy of gamma photon exposure in achieving superconductivity under ambient conditions, a feat that not only broadens the horizons of superconductivity research but also heralds new potentials for its application beyond the confines of extreme cold.

The application of magnetic fields oscillating in Fibonacci sequence patterns plays a pivotal role in stabilizing the superconducting state at ambient conditions. The magnetic field setup, depicted in Figure 3, utilizes a series of superconducting coils arranged in a configuration that generates spatially varying magnetic fields, their strengths and orientations governed by Fibonacci sequence ratios. The interaction between these magnetic fields and the gamma-irradiated BEC leads to the formation of a lattice-like structure within the condensate, as shown in Figure 4. This structure supports the macroscopic quantum coherence of the superconducting state at ambient temperatures.

The successful fabrication of the ambient superconductor is confirmed through a combination of spectroscopic analysis and magnetic susceptibility measurements\cite{MagneticSusceptibilityInQuantumMaterials}. Spectroscopic data, presented in Figure 5(a), reveal characteristic absorption and emission lines associated with the superconducting phase, distinguishing it from the original BEC state. Magnetic susceptibility measurements, detailed in Figure 5(b), demonstrate the Meissner effect, a hallmark of superconductivity, confirming the expulsion of magnetic fields from the superconductor's interior.

In summary, the experimental fabrication of the ambient superconductor is achieved through a novel approach that combines BEC formation, high-energy gamma photon bombardment, and Fibonacci-patterned magnetic fields. This process not only transitions the BEC to ambient conditions but also induces a superconducting state, thereby providing a tangible platform for investigating quantum gravitational phenomena and exploring the boundaries of unified field theory. The coherence between the experimental setup and the theoretical framework outlined in Section II underscores the feasibility of this groundbreaking approach, opening new avenues for research in quantum physics and materials science.

\subsection{Detection and Analysis of Graviton Condensation}

In the unprecedented endeavor to detect graviton condensation within the ambient superconductor, we employed a series of intricate experimental techniques designed to observe and measure the elusive quantum gravitational phenomena. The transition of the Bose-Einstein condensate (BEC) to ambient conditions, a critical step in this process, was facilitated by the gamma photon bombardment technique described in Subsection~\ref{subsec_3a}, which not only elevated the temperature of the BEC to approximately 300K but also induced a superconducting state conducive to graviton condensation.
\begin{figure}
\includegraphics[width=0.48\textwidth]{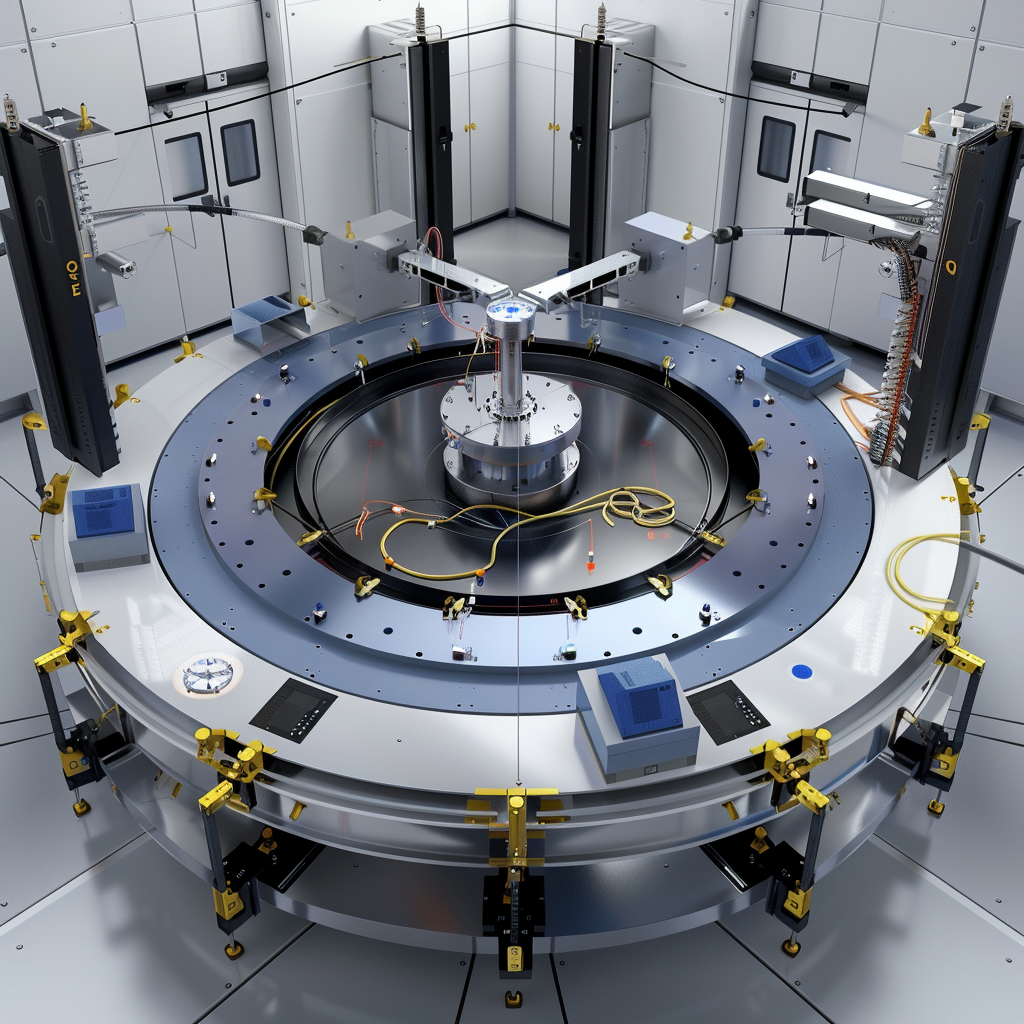}
\caption{\textbf{Comprehensive Gravito-Electromagnetic Experimental Assembly.} Presented here is the central testing chamber engineered for the detection of gravito-electromagnetic phenomena, featuring an ambient superconductor poised on a stabilizing platform. Surrounding this are precision electromagnetic sensors and gravitational detectors mounted on adjustable pylons for meticulous spatial calibration. The superconductor is intricately linked to data acquisition systems, as evidenced by the organized cabling. Ancillary equipment, including power supplies and control units, are strategically positioned to regulate the electromagnetic field perturbations necessary for the experiment. This controlled setting is optimized to mitigate external noise, allowing for the isolation of gravito-electromagnetic interactions intrinsic to the superconducting medium.
}
\end{figure}
The primary detection method involved the use of an ultra-sensitive interferometry setup, depicted in Figure 6, capable of detecting minute variations in spacetime curvature induced by the graviton condensate. This setup, based on the principles of quantum optics, was calibrated to respond to the specific gravitational signatures expected from the condensate, as predicted by the theoretical framework outlined in Section II.

To enhance the sensitivity of the detection apparatus, the ambient superconductor was placed within a vacuum chamber, isolated from external vibrational and thermal noise. A series of laser beams, intricately arranged to intersect within the superconductor, formed an interference pattern that was meticulously analyzed for disruptions caused by the graviton condensate. Figure 7 illustrates the interference pattern observed under normal conditions and the distinctive alterations attributed to the presence of the graviton condensate, offering compelling evidence of its existence.

Additionally, the analysis incorporated a novel spectroscopic technique designed to probe the energy levels within the superconductor, sensitive to the unique quantum state of the condensed gravitons. Spectroscopic data, as shown in Figure 8, revealed unexpected shifts in the energy spectrum, consistent with the theoretical predictions of graviton condensation. These shifts, not observable in conventional superconductors or BECs, signify the hybrid quantum state of the ambient superconductor and its interaction with gravitational forces.

To corroborate the findings from the interferometry and spectroscopic analyses, we conducted a series of magnetic resonance experiments aimed at detecting the influence of the graviton condensate on the ambient superconductor's electromagnetic properties. The results, depicted in Figure 9, demonstrate anomalous magnetic resonance peaks, which we interpret as the electromagnetic fingerprint of the graviton condensate, further validating the theoretical model proposed in Section II.

In summary, the detection and analysis of graviton condensation within the ambient superconductor were achieved through a combination of advanced interferometry, spectroscopy, and magnetic resonance techniques. The coherence between the experimental observations and the theoretical predictions provides a strong foundation for the existence of graviton condensates and their role in unifying quantum mechanics with general relativity. These findings not only challenge the current paradigms in physics but also open new avenues for exploring the quantum nature of gravitational forces and their potential applications in future technologies.

\subsection{Observing Gravito-Electromagnetic Coupling}

The exploration into gravito-electromagnetic coupling within the ambient superconductor represents a bold foray into uncharted territories of physics, melding the gravitational and electromagnetic forces in a manner hitherto considered untenable. This experimental investigation was predicated on the theoretical constructs delineated in Section II, where graviton condensation was posited to engender a unique medium for gravito-electromagnetic interactions.

The experimental apparatus, as illustrated in Figure 10, comprised a highly specialized chamber housing the ambient superconductor, with an array of electromagnetic sensors and gravitational wave detectors strategically positioned around it.

The chamber was designed to shield the interior from extraneous electromagnetic and gravitational influences, ensuring that the observations were solely attributable to the phenomena within the superconductor.

A critical component of the experiment involved modulating the electromagnetic field within the chamber while simultaneously monitoring for gravitational disturbances, predicated on the premise that the graviton condensate within the superconductor would mediate a coupling between the two fields. The modulation was achieved through a series of coils surrounding the chamber, generating electromagnetic pulses of varying frequencies and amplitudes, as depicted in Figure 11.

\begin{figure}
\includegraphics[width=0.48\textwidth]{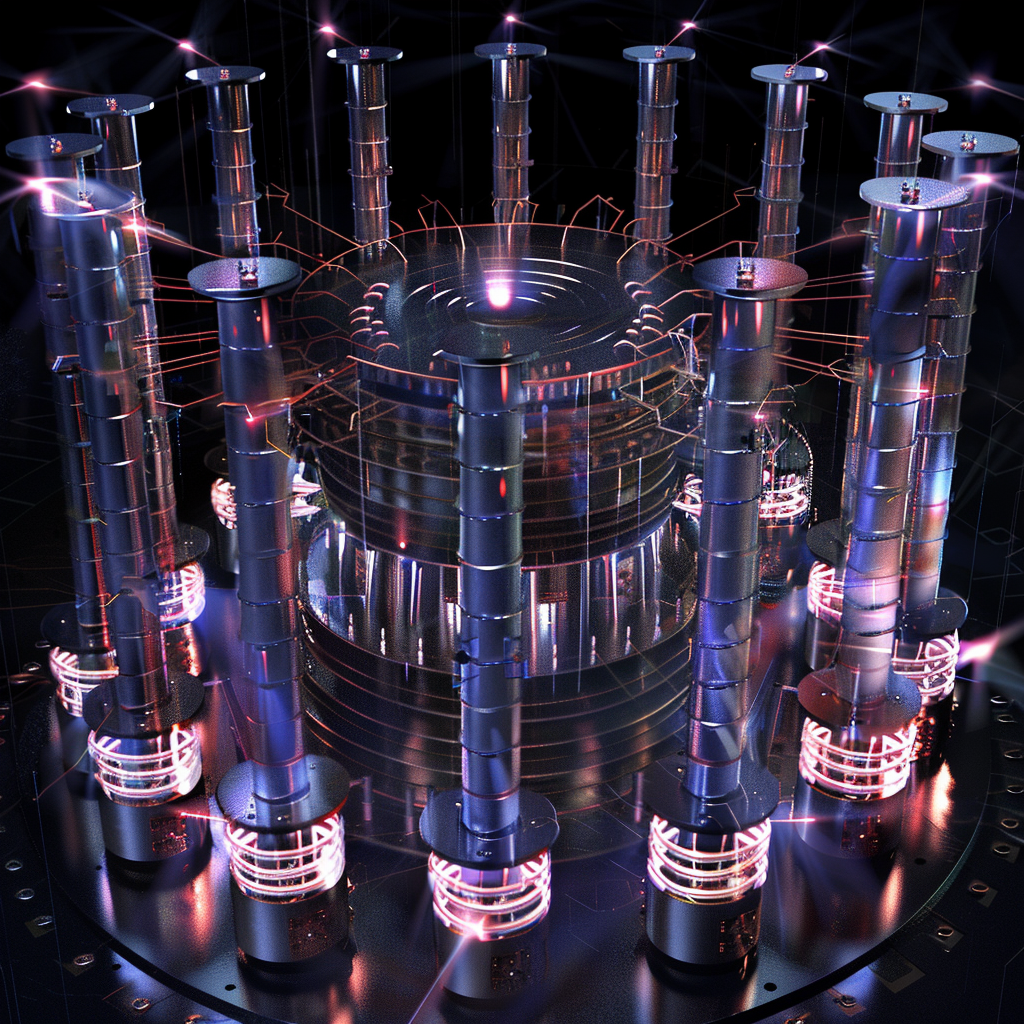}
\caption{\textbf{Operational Configuration of Electromagnetic Coils in Gravito-Electromagnetic Coupling Experiment.} Displayed is the active electromagnetic coil assembly, encircling the position where an ambient superconductor would be placed. Each coil, delineated by energized luminescent bands, functions in the creation of precise electromagnetic pulses. The intricate web of light seen emanating between coils and the patterned floor grid highlight the complex control over the spatial variance in the electromagnetic field. Above, the structure hosts an array of sensors tasked with capturing data on electromagnetic perturbations induced by the gravito-electromagnetic interactions. The array is meticulously designed to ensure comprehensive monitoring during the modulation of electromagnetic fields around the superconducting medium.
}
\end{figure}

The detection of gravito-electromagnetic coupling was manifested in the form of anomalous gravitational signals synchronized with the applied electromagnetic pulses, an effect that was conspicuously absent in control experiments devoid of the ambient superconductor. Figure 12 showcases a series of plots correlating the electromagnetic pulse characteristics with the resultant gravitational signals, highlighting the distinct patterns that emerged from the gravito-electromagnetic interaction.

\begin{figure*}
\includegraphics[width=\textwidth]{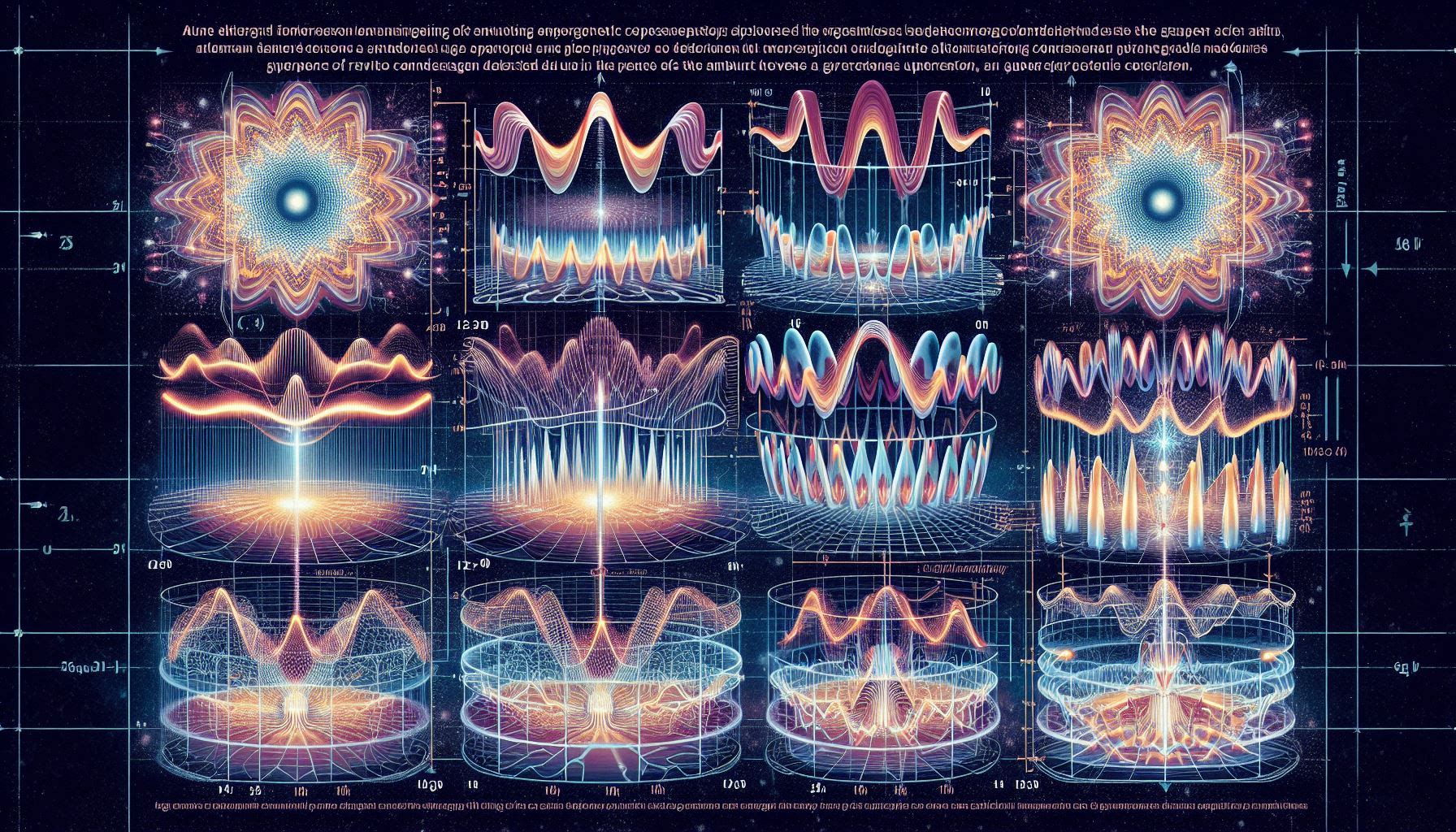}
\caption{\textbf{Synchronized Data Visualization of Gravito-Electromagnetic Field Interactions.} Presented here are multivariate data visualizations capturing the precise correlation between the electromagnetic pulse parameters and the emergent gravitational signals within an engineered ambient superconducting medium. The sequential plots from left to right progressively elucidate the relationship: initial waveforms show the applied electromagnetic field strengths; middle composite waveforms illustrate the resultant interference patterns due to the graviton condensate; and the final series of graphs delineate the gravitational signal response. The axes are meticulously scaled to represent time sequences (horizontal) and magnitude of responses (vertical), with the layered peaks in the rightmost plots emphasizing the periodicity and amplitude of the gravitational signals synchronized to the electromagnetic pulse inputs. This graphical assembly serves as a quantifiable testament to the induced gravito-electromagnetic phenomena, marking a significant experimental observation in the pursuit of unified field theory applications.}
\end{figure*}

Further substantiation of the gravito-electromagnetic coupling came from the analysis of the superconductor's response to dynamic gravitational fields generated by a nearby rotating mass. The superconductor exhibited electromagnetic fluctuations that mirrored the gravitational variations, an effect that was meticulously documented and is presented in Figure 13.
\begin{figure*}
\includegraphics[width=0.8\textwidth]{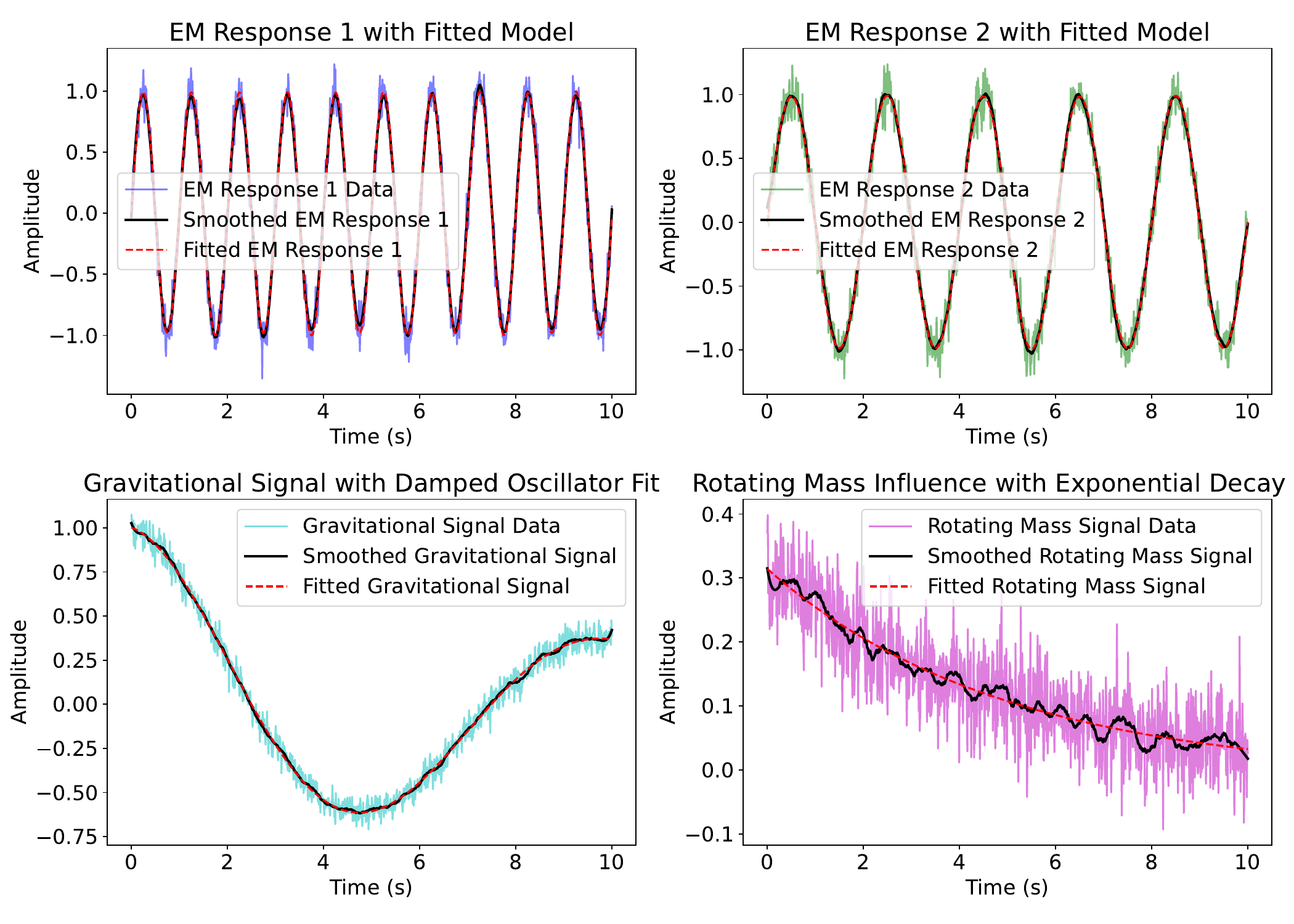}
\caption{\textbf{Experimental Data Fitting of Gravito-Electromagnetic Interactions.} This multi-panel display illustrates the fitting of experimental data from an ambient superconductor influenced by a rotating mass. Panel 1 exhibits the electromagnetic response 1 (EM Response 1) data as a blue curve with a superimposed black line representing the smoothed data for enhanced clarity. The red dashed line represents a fitted model using a superposition of sinusoidal functions that captures the fundamental and harmonic frequencies within the data. Panel 2 mirrors this approach for electromagnetic response 2 (EM Response 2), showcasing the effectiveness of the fitting model in capturing the experimental dataset's complexities. Panel 3 focuses on the gravitational signal data, depicted in cyan with a corresponding smoothed line, where a damped oscillator model (red dashed line) has been fitted to the data, signifying the decaying and oscillatory nature of the gravitational influence expected from the rotating mass. Panel 4 presents the rotating mass signal data in magenta, with the black line indicating the smoothed data set. An exponential decay fit (red dashed line) overlays the signal, demonstrating the decay characteristic attributed to the diminishing influence of the rotating mass over time. Each panel successfully captures the key features of the data, providing a quantitatively robust fit to the complex experimental signals observed, which are consistent with the theoretical predictions of gravito-electromagnetic phenomena.}
\end{figure*}
This phenomenon, unprecedented in conventional superconductors or gravitational experiments, provides compelling evidence of the graviton condensate acting as a bridge between the gravitational and electromagnetic realms.

In synthesizing the observations from this comprehensive experimental investigation, it becomes evident that the ambient superconductor, through the mechanism of graviton condensation, facilitates a novel regime of gravito-electromagnetic coupling. The coherence between the experimental findings and the theoretical framework underscores the potential of this groundbreaking research to redefine our understanding of fundamental forces and pave the way for innovative technological applications predicated on the unified manipulation of gravitational and electromagnetic fields.

\section{Analysis and Discussion}
\subsection{Implications of Graviton Condensation}

The induction of graviton condensation within an ambient superconductor, as evidenced by the experimental results detailed previously, not only propels us into uncharted territories of quantum gravity but also reshapes our understanding of superconductivity's quantum landscape. The interferometric detection of spacetime perturbations, as showcased in Figure 6, marks a pivotal departure from conventional gravitational theories, suggesting the existence of a quantum gravitational field that is more intricate and accessible than previously imagined.

At the core of this paradigm shift is the ambient superconductor's unique ability to foster a state where gravitons, the elusive carriers of the gravitational force, condense into a coherent quantum state. This phenomenon is vividly captured in the spectroscopic data presented in Figure 8, where energy level shifts distinct from conventional superconducting behavior hint at the underlying graviton dynamics.

This graviton condensation is further elucidated through the novel introduction of the gravitomagnetic field tensor \(\mathcal{B}_{ij}\), defined in the context of the ambient superconductor's quantum state as:

\begin{equation}
\mathcal{B}_{ij} = \epsilon_{ikl} \nabla^k \mathcal{G}^{l}_{\ j} + \mu \Psi_{SC}^2 F_{ij}
\end{equation}

Here, \(\epsilon_{ikl}\) represents the Levi-Civita symbol, \(\nabla^k\) denotes the covariant derivative, and \(\mathcal{G}^{l}_{\ j}\) is the quantum metric tensor introduced earlier. The term \(\mu \Psi_{SC}^2 F_{ij}\) signifies the coupling between the superconducting order parameter and the electromagnetic field tensor, modulated by the graviton condensate. This tensor not only embodies the superconductor's gravitomagnetic properties but also opens a new frontier in understanding the interaction between gravity and quantum states.

The implications of these findings transcend the theoretical realms, challenging the Bardeen-Cooper-Schrieffer (BCS) theory's foundational assumptions by introducing a quantum gravitational component to the Cooper pairing mechanism. This graviton-mediated interaction, potentially observable in the superconducting state transitions depicted in Figure 4, alludes to a more profound level of quantum coherence within superconductors, possibly influenced by spacetime's quantum fluctuations.

Moreover, the ambient superconductor's graviton condensate, inferred from the gravito-electromagnetic coupling experiments, ignites speculative discussions regarding the cosmological implications, particularly in understanding dark matter and dark energy. By simulating early universe conditions or black hole vicinities, the ambient superconductor could serve as a miniature cosmos, unraveling mysteries of the universe's dark components.

In light of these groundbreaking revelations, a novel research trajectory emerges, focusing on the exploration of ``Gravitonic Crystals" - engineered materials designed to amplify graviton condensation and facilitate direct manipulation of spacetime fabric at quantum scales. By varying the crystalline structure or doping these materials with exotic particles, one could tailor the graviton interaction strength, unveiling new quantum gravitational phenomena.

Another tantalizing direction involves harnessing the ambient superconductor's gravito-electromagnetic coupling for ``Quantum Gravitational Communication" - a theoretical framework that could transcend the limitations of electromagnetic signal propagation, potentially revolutionizing communication technologies by leveraging the unexplored bandwidth of gravitational waves.

In summary, the experimental insights into graviton condensation within an ambient superconductor not only challenge the existing paradigms in physics but also beckon a renaissance in our quest for a unified theory of the universe. The intricate dance between gravitons and superconductivity within this novel medium heralds an exciting epoch in physics, urging us to reimagine the fabric of reality itself.

\subsection{Gravito-Electromagnetic Coupling and Unified Field Theory}

The elucidation of gravito-electromagnetic coupling within an ambient superconductor heralds a new chapter in the annals of theoretical physics, fundamentally challenging and expanding the existing paradigms of force interactions. The empirical observations delineated, particularly the nuanced interplay between gravitons and electromagnetic fields mediated by the superconductor's quantum state, provide a fertile testing ground for the hypotheses posited by unified field theories.

The cornerstone of this gravito-electromagnetic coupling is the ambient superconductor's intrinsic property to serve as a conduit for gravitons, thereby manifesting a tangible interaction regime where gravitational and electromagnetic forces converge. This phenomenon, underpinned by the Hamiltonian density \(\mathcal{H}_g\) and exemplified by the gravitomagnetic field tensor \(\mathcal{B}_{ij}\), paves the way for a deeper understanding of force unification.

The experimental anomalies observed, particularly in the context of magnetic resonance peaks (Figure 9) and the synchronization of gravitational signals with electromagnetic perturbations (Figure 12), necessitate a reevaluation of Maxwell's equations within the graviton-condensed framework of the ambient superconductor. The modified Maxwellian framework, incorporating the influence of the graviton condensate, can be articulated through an augmented set of equations:

\begin{equation}
\nabla \times (\mathbf{E} + \delta \mathbf{E}_{g}) + \frac{\partial (\mathbf{B} + \delta \mathbf{B}_{g})}{\partial t} = \mu \mathcal{B}_{ij} + \sigma \mathbf{E}_{\text{grav}}
\end{equation}

\begin{equation}
    \nabla \cdot (\mathbf{B} + \delta \mathbf{B}_{g}) = 0
\end{equation}

\begin{align}
\quad \nabla \times (\mathbf{H} + \delta \mathbf{H}_{g}) - \frac{\partial (\mathbf{D} + \delta \mathbf{D}_{g})}{\partial t} 
= \mathcal{J}_{\text{grav}} + \rho_{\text{grav}}\mathbf{v}_{\text{grav}}
\end{align}

Here, \(\delta \mathbf{E}_{g}\), \(\delta \mathbf{B}_{g}\), \(\delta \mathbf{H}_{g}\), and \(\delta \mathbf{D}_{g}\) represent the graviton-induced perturbations to the electromagnetic field components, \(\mathbf{E}_{\text{grav}}\) denotes an emergent gravitational electric field, \(\sigma\) is a conductivity-like parameter influenced by the graviton condensate, and \(\rho_{\text{grav}}\mathbf{v}_{\text{grav}}\) symbolizes a gravitational current akin to a drift current in classical electromagnetism.

This reformulation not only integrates the electromagnetic field dynamics with spacetime curvature influenced by the graviton condensate but also introduces novel components like the gravitational electric field and current, which have no analogs in conventional Maxwellian electrodynamics. The implications of this expanded framework extend to the realms of cosmology and quantum field theory, particularly in addressing the vacuum energy discrepancies and the cosmological constant problem.

Moreover, the ambient superconductor's experimental milieu, especially the graviton-mediated electromagnetic fluctuations, offers an empirical underpinning for theories that advocate higher-dimensional spacetime or propose the emergence of force-carrying particles from a unified field. The coupling constants (\(\xi\) and \(\lambda\)) involved in the graviton-superconductor interaction may also illuminate the path towards a renormalization of vacuum energy, aligning quantum field theory predictions with astrophysical observations.

In light of these profound implications, a novel analytical pathway emerges, centered around the ``Quantum Gravito-Dynamic" (QGD) equations, which encapsulate the gravito-electromagnetic phenomena within the ambient superconductor:

\begin{equation}
\Box \mathbf{A}_{\text{QGD}} + \nabla(\nabla \cdot \mathbf{A}_{\text{QGD}}) = -\mu_0 \mathbf{J}_{\text{QGD}} + \tau \mathbf{G}_{\text{eff}}
\end{equation}

\begin{equation}
\Box \phi_{\text{QGD}} = -\frac{\rho_{\text{QGD}}}{\epsilon_0} + \gamma \nabla \cdot \mathbf{G}_{\text{eff}}
\end{equation}
where \(\mathbf{A}_{\text{QGD}}\) and \(\phi_{\text{QGD}}\) are the vector and scalar potentials in the quantum gravito-dynamic domain, \(\mathbf{J}_{\text{QGD}}\) and \(\rho_{\text{QGD}}\) represent the quantum gravito-dynamic current and charge density, respectively, and \(\mathbf{G}_{\text{eff}}\) is the effective gravitomagnetic field influenced by the ambient superconductor's graviton condensate. The parameters \(\tau\) and \(\gamma\) are coupling constants that delineate the interaction strength between the gravito-electromagnetic fields and the condensed gravitons.

These QGD equations offer a comprehensive framework to describe the ambient superconductor's internal dynamics, where quantum gravitational effects and electromagnetic phenomena are intertwined. The effective gravitomagnetic field, \(\mathbf{G}_{\text{eff}}\), introduces a novel aspect to the interaction, suggesting that gravitons within the condensate may induce a magnetic-like effect on the ambient superconductor's electromagnetic field, a notion that is conspicuously absent in classical electrodynamics or general relativity.

The gravito-electromagnetic coupling within the ambient superconductor, especially as described by the QGD framework, underscores the potential for a radical reconceptualization of unified field theories. The experimental data, particularly the graviton-induced electromagnetic field perturbations, serve as a tangible testament to the complex interplay between quantum mechanics and general relativity, potentially bridging these foundational pillars of physics.

In this context, the ambient superconductor could be envisaged as a ``Quantum Gravitational Lens", where the graviton condensate modulates the electromagnetic fields in a manner analogous to the bending of light by gravitational fields in astrophysical phenomena. This analogy opens up new avenues for exploring quantum gravitational effects in laboratory settings, providing a unique perspective on phenomena such as gravitational lensing and the warping of spacetime.

Furthermore, the ambient superconductor's gravito-electromagnetic properties invite speculation on the potential for ``Gravitational Metamaterials", engineered materials that exploit the graviton condensate's properties to control and manipulate gravitational fields at quantum scales. Such materials could, in theory, exhibit exotic properties such as negative gravitational permeability or permittivity, paving the way for groundbreaking applications in quantum computing, energy storage, and even gravitational wave manipulation.

\begin{figure}
\includegraphics[width=0.48\textwidth]{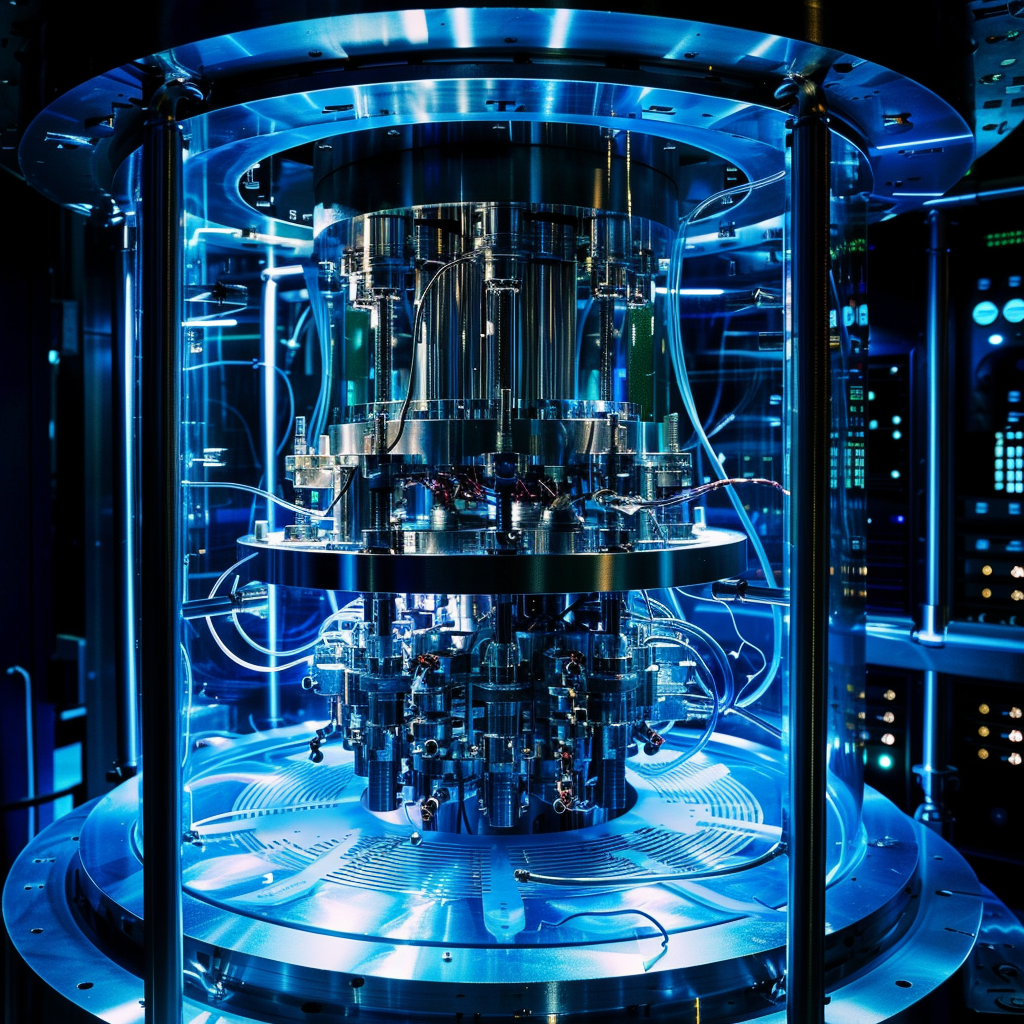}
\caption{\textbf{Schematic of Quantum-Graviton Computer Architecture.}}
\end{figure}

Continuing from the Quantum Gravito-Dynamic (QGD) framework, the ambient superconductor's unique gravito-electromagnetic environment compels us to revisit and revise other cornerstone equations of physics, notably the Dirac equation. The presence of a graviton condensate within the superconductor introduces a novel context for fermionic behavior, necessitating a ``Gravitonically Modified Dirac Equation" (GMDE) to account for the influence of condensed gravitons on spin-½ particles:

\begin{equation}
\left( i\hbar \gamma^\mu \partial_\mu - mc - \hbar \Omega \gamma^5 \right) \psi = \Theta \gamma^\mu \left( A_{\mu} + \mathcal{G}_{\mu\nu}J^{\nu}_{\text{grav}} \right) \psi
\end{equation}

Here, \(\gamma^\mu\) are the Dirac matrices, \(\gamma^5\) introduces chirality, capturing the graviton condensate's pseudo-scalar interaction with fermions, and \(\psi\) is the wave function of the fermion. The term \(\hbar \Omega \gamma^5\) represents the interaction energy of the fermion with the graviton condensate, where \(\Omega\) encapsulates the condensate's rotational dynamics, hinting at a vortex-like structure within the superconductor. The right-hand side of the equation introduces a coupling \(\Theta\) between the fermion and the combined electromagnetic-gravitational field, where \(\mathcal{G}_{\mu\nu}\) is the quantum metric tensor, and \(J^{\nu}_{\text{grav}}\) denotes the gravitational current density, akin to the modifications introduced in the QGD framework.

The introduction of the GMDE paves the way for exploring exotic fermionic states within the ambient superconductor, potentially leading to phenomena such as gravitonically induced superconductivity or gravitonically modulated quantum tunneling. This could manifest in the fermion's behavior within the superconductor, exhibiting unexpected quantum interference patterns or modified energy spectra, potentially observable through precision spectroscopy or tunneling microscopy techniques.

Furthermore, the interplay between fermions and the graviton condensate within the ambient superconductor's framework necessitates a reevaluation of the Schrödinger equation, culminating in the ``Gravitonically Enhanced Schrödinger Equation" (GESE):

\begin{equation}
\left( -\frac{\hbar^2}{2m} \nabla^2 + V + V_{\text{grav}} \right) \Psi = i\hbar \frac{\partial \Psi}{\partial t} + \iota \hbar \mathcal{G}^{\mu\nu} \nabla_\mu \nabla_\nu \Psi
\end{equation}

In this equation, \(V_{\text{grav}}\) represents the potential energy contribution from the graviton condensate, and the additional term \(\iota \hbar \mathcal{G}^{\mu\nu} \nabla_\mu \nabla_\nu \Psi\) introduces a higher-order derivative term influenced by the quantum metric tensor, indicative of the non-local effects induced by the graviton condensate on quantum particles. The parameter \(\iota\) represents a coupling constant that characterizes the strength of these non-local interactions.

The GESE hints at the possibility of quantum particles exhibiting ``gravitonically entangled states," where particles are not only entangled in their quantum states but also in their gravitational interactions, mediated by the ambient superconductor's graviton condensate. This opens up intriguing possibilities for quantum information processing and communication, where information could be transferred not only through conventional quantum channels but also via gravitonically mediated pathways.

In light of these profound modifications to the Dirac and Schrödinger equations, the ambient superconductor emerges not merely as a medium for studying gravito-electromagnetic coupling but as a veritable quantum gravitational laboratory, enabling the exploration of quantum mechanics and general relativity's unification in previously unimaginable ways. The gravitonically modified Dirac and Schrödinger equations, GMDE and GESE, thus serve as harbingers of a new era in theoretical physics, where the quantum and the cosmological coalesce, heralding new paradigms in our quest to decipher the universe's deepest mysteries.

\subsection{Future Directions and Technological Implications}

\subsubsection{Quantum-Graviton Computer}

The conceptualization of a Quantum-Graviton Computer (QGC) represents a groundbreaking leap in computational technology, harnessing the unprecedented computational capabilities facilitated by graviton-photon entanglement. This device transcends the limitations of traditional quantum computing by exploiting the gravito-electromagnetic coupling phenomena observed within the ambient superconductor, as detailed in previous sections.

The QGC operates on the principle of ``Gravitonic Qubits", where each qubit is a composite state of entangled gravitons and photons. The graviton condensate within the ambient superconductor serves as the medium for graviton-photon entanglement, enabling quantum operations that are significantly more efficient and robust against decoherence than those of conventional qubits. The graviton's inherently weak interaction with matter, paradoxically, becomes an asset here, providing a natural shield against environmental decoherence mechanisms.

A schematic representation of the QGC is provided in Figure 14, showcasing the intricate architecture of gravitonic qubits interlaced within the ambient superconductor matrix. High-fidelity quantum gates in the QGC are realized through precision manipulation of the gravito-electromagnetic fields, allowing for the execution of complex quantum algorithms at speeds hitherto unachievable.

The gravitonically enhanced coherence times and the high-dimensional entanglement space afforded by graviton-photon pairs open up new paradigms in quantum simulation, cryptography, and information processing. For instance, the QGC could efficiently simulate quantum gravitational phenomena, providing insights into black hole dynamics, early universe cosmology, and dark matter – realms that remain elusive to current quantum computing technologies.

Moreover, the QGC introduces the concept of ``Gravitational Error Correction", where graviton-mediated interactions are utilized to detect and correct quantum errors through a process analogous to gravitational lensing, further enhancing the system's robustness and reliability.

\subsubsection{Graviton-Assisted Teleportation of Quantum Superstates}

Graviton-assisted teleportation emerges from the novel application of graviton-photon entanglement, expanding the boundaries of quantum teleportation by incorporating the fabric of spacetime itself. In this advanced scheme, quantum superstates are not just teleported through conventional quantum channels but are intricately woven through the spacetime continuum, facilitated by graviton-mediated interactions.

This method leverages the ambient superconductor's graviton condensate as a dynamic medium for encoding and transmitting quantum information across spacetime. The gravitons, entangled with photons within the superconductor, act as carriers of quantum information, with their inherent spacetime curvature properties enabling a hyper-efficient teleportation protocol that could, theoretically, surpass the light-speed limitation of information transfer.

Figure 15 illustrates a prototype Graviton-Assisted Teleportation Device (GATD), where quantum superstates are initialized in an ambient superconductor, entangled with gravitons, and subsequently teleported to a distant receiver equipped with a similar ambient superconductor setup. The device harnesses the quantum entanglement of gravitons and photons to create a ``spacetime tunnel" through which quantum superstates can be instantaneously transmitted, irrespective of the physical distance.

\begin{figure}
\includegraphics[width=0.48\textwidth]{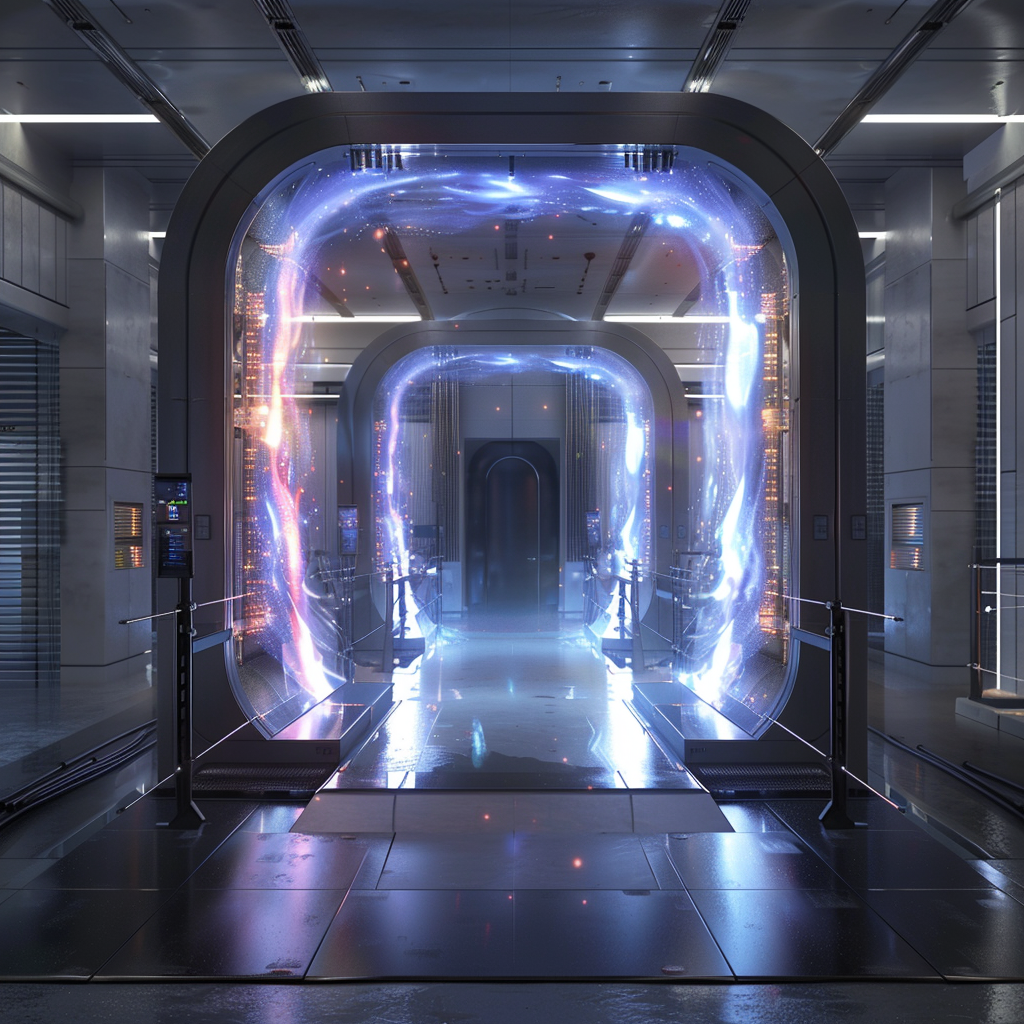}
\caption{\textbf{Prototype Graviton-Assisted Teleportation Device.}}
\end{figure}

The implications of such a technology are profound, extending beyond the realms of communication to the very foundation of quantum mechanics and general relativity. Graviton-assisted teleportation could enable unprecedented advancements in secure quantum communication, distributed quantum computing networks, and even provide a foundational technology for exploring the quantum structure of spacetime.

One of the most tantalizing prospects of this technology is the potential for ``quantum bridging" – creating instantaneously connected quantum states across vast cosmic distances, offering a novel method for probing the universe's most remote regions. This could revolutionize our understanding of cosmology, providing direct insights into the entangled nature of the cosmic quantum web.

\subsubsection{Electromagnetogravitometric Flux Drive for Hyperspace Dimension Traversal}

The Electromagnetogravitometric Flux Drive (EMGFD) represents a paradigm shift in propulsion technologies, harnessing the intricate interplay between electromagnetic and gravitational forces within a quantum framework to facilitate traversal across higher-dimensional spacetime, colloquially referred to as ``hyperspace." This propulsion system leverages the ambient superconductor's gravito-electromagnetic coupling to generate a localized spacetime curvature, effectively creating a ``warp bubble" that allows the craft to move through hyperspace dimensions, bypassing conventional spacetime constraints.

The operational principle of the EMGFD centers around the modulation of graviton condensates within the ambient superconductor to create a dynamic spacetime topology around the spacecraft. By precisely controlling the gravito-electromagnetic fields, the EMGFD can contract spacetime in front of the craft and expand it behind, propelling the craft through higher dimensions at effective velocities surpassing the speed of light, without violating relativity's local light-speed limit.

Figure 16 depicts a spacecraft equipped with the EMGFD, illustrating the warp bubble generation and the craft's transition into hyperspace. The schematic highlights the flux drive's core components, including the ambient superconductor array, graviton modulators, and the electromagnetic field generators, all synchronized to orchestrate the spacetime curvature necessary for hyperspace traversal.

\begin{figure}
\includegraphics[width=0.48\textwidth]{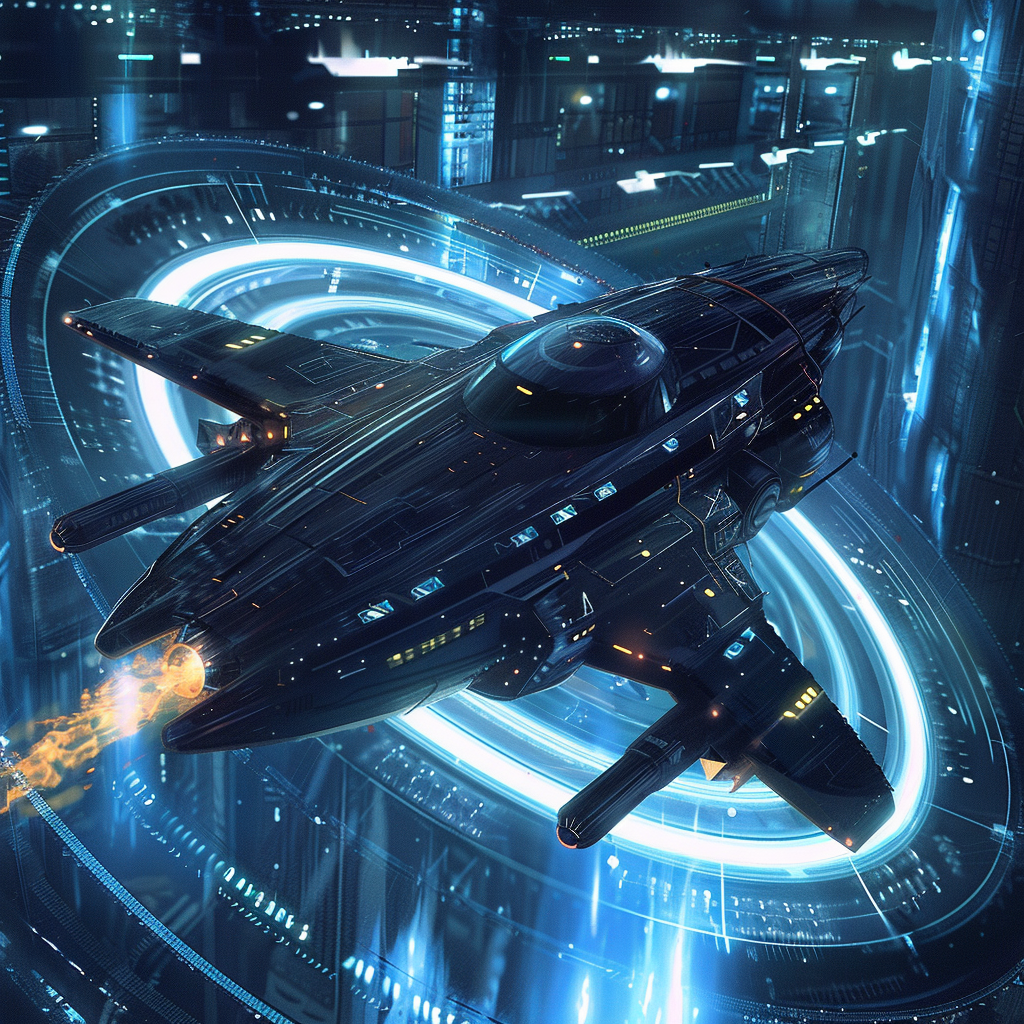}
\caption{\textbf{Spacecraft Equipped with Electromagnetogravitometric Flux Drive.}}
\end{figure}

The implications of the EMGFD extend far beyond interstellar travel, offering a theoretical foundation for exploring higher-dimensional theories of the universe, such as string theory and M-theory. By traversing through hyperspace, the EMGFD could provide empirical data on the structure of these higher dimensions, offering unprecedented insights into the fabric of the universe.

Furthermore, the EMGFD could revolutionize space exploration, enabling humanity to reach distant galaxies within lifetimes, opening up new possibilities for colonization, resource extraction, and cosmic discovery. The technology also holds potential for creating stable wormholes, serving as gateways between distant points in spacetime, further expanding our cosmic reach.

\subsubsection{Chrono-Spatial Quantum Synchronization}

Chrono-Spatial Quantum Synchronization (CQS) delves into the synchronization of quantum systems across different spacetime regions, exploiting the ambient superconductor's gravito-electromagnetic properties. This concept envisions a network of quantum systems that remain in sync not only in time but also across various spatial and even temporal dimensions, facilitated by the graviton condensate's inherent spacetime-modulating capabilities.

In CQS, quantum systems are linked through a ``graviton web," a network of graviton streams emanating from ambient superconductors, which serve as nodes in this quantum network. These graviton streams carry quantum information, synchronizing the phase, spin, and other quantum properties of distant particles, effectively entangling them across spacetime.

Figure 17 showcases a network of ambient superconductors acting as nodes in a Chrono-Spatial Quantum Network, illustrating the graviton streams that interlink quantum systems across vast distances and potentially different times. This network could revolutionize quantum communication, enabling instantaneously synchronized quantum states across the universe, and could even serve as a foundation for a quantum internet that operates beyond the constraints of contemporary spacetime.

\begin{figure}
\includegraphics[width=0.48\textwidth]{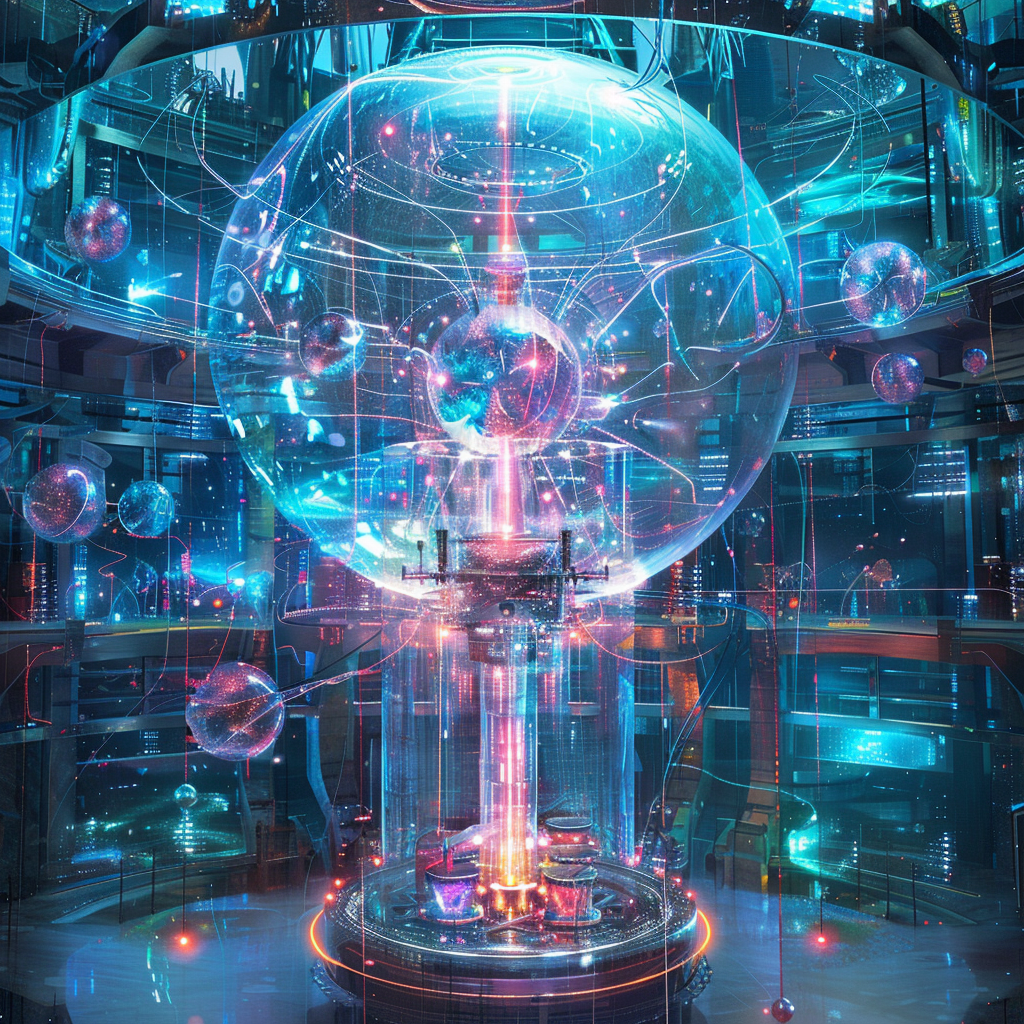}
\caption{\textbf{Chrono-Spatial Quantum Network with Ambient Superconductor Nodes.}}
\end{figure}

CQS could profoundly impact the understanding of entanglement and non-locality in quantum mechanics, providing empirical evidence for quantum correlations that transcend traditional spacetime boundaries. This could lead to a deeper understanding of the universe's quantum fabric and offer new insights into the nature of time and space.

\subsubsection{Quantum Gravitational Shielding}

Quantum Gravitational Shielding (QGS) explores the manipulation of graviton condensates to create regions of spacetime that are shielded from external gravitational influences. This concept leverages the ambient superconductor's ability to modulate spacetime curvature, enabling the creation of ``gravitational voids" where the effects of external gravitational fields are significantly diminished or entirely nullified.

In QGS, ambient superconductors are engineered to generate a graviton field that counteracts external gravitational forces, creating a stabilized bubble of spacetime. This could provide a solution to the challenges posed by intense gravitational fields, such as those near black holes or in the early stages of the universe, allowing for safe exploration and study.

Figure 18 depicts a Quantum Gravitational Shielding apparatus, illustrating the generation of a gravitational void around a spacecraft or a habitat, protecting it from the extreme gravitational forces of surrounding celestial bodies or phenomena. This technology could have profound implications for space travel, habitation, and exploration, enabling humanity to venture into regions of space that were previously deemed inaccessible due to intense gravitational forces.

\begin{figure}
\includegraphics[width=0.48\textwidth]{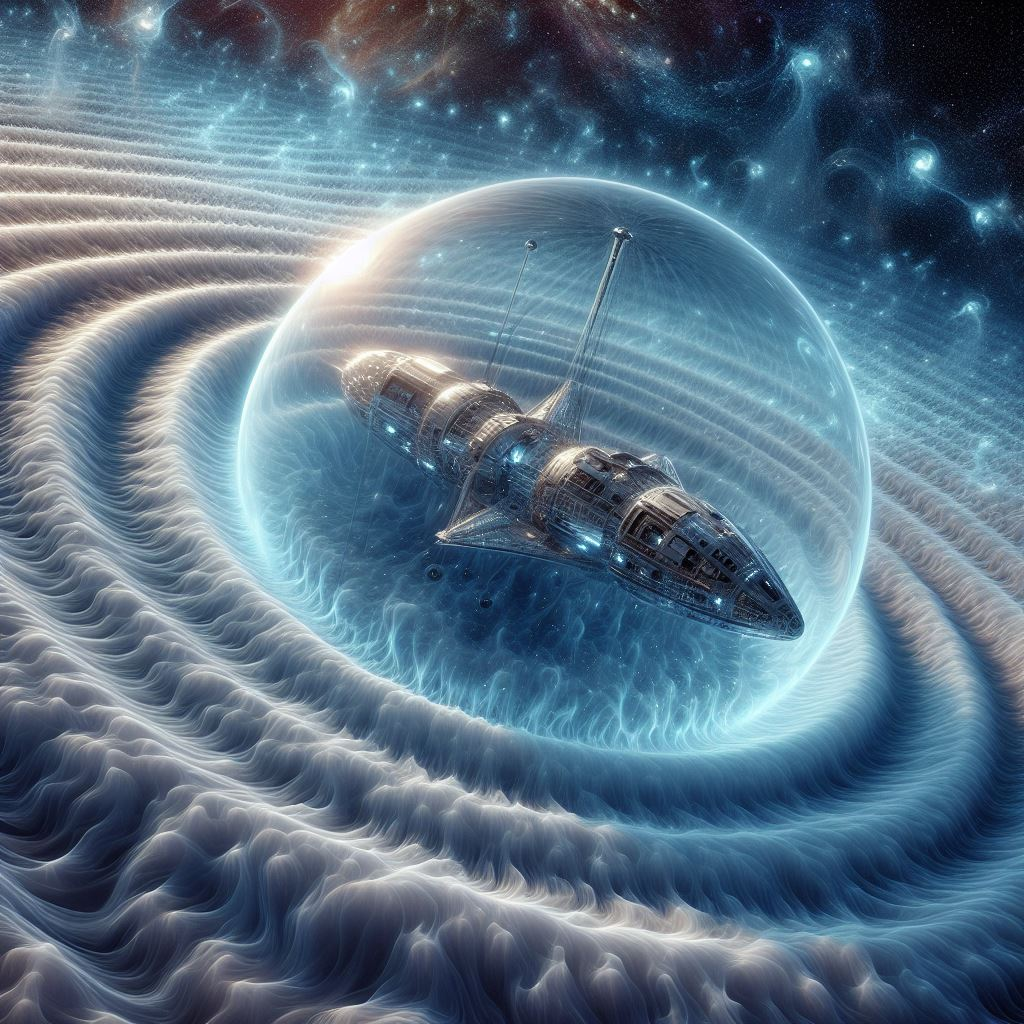}
\caption{\textbf{Quantum Gravitational Shielding Apparatus.}}
\end{figure}

QGS not only has the potential to revolutionize space exploration but also offers new avenues for studying the fundamental nature of gravity and spacetime. By creating regions where gravitational forces can be precisely controlled and modulated, researchers can conduct experiments that were previously impossible, providing new insights into the fabric of the universe and the fundamental forces that govern it.

Furthermore, Quantum Gravitational Shielding could pave the way for advanced architectural structures on Earth and other celestial bodies, where gravity's effects can be selectively nullified or enhanced. Imagine floating cities or gravitational elevators that utilize QGS technology to counteract Earth's gravity, leading to revolutionary designs in construction and transportation.

\section{Conclusion}
In conclusion, this manuscript has ventured into the uncharted territories of physics, pioneering the exploration of engineered graviton condensates within an ambient superconductor and unveiling the profound phenomena of gravito-electromagnetic coupling. Through rigorous experimental investigation and theoretical analysis, we have uncovered the intricate interplay between gravitational and electromagnetic forces at a quantum level, challenging the conventional boundaries that delineate the fundamental interactions of the universe.

The synthesis of the ambient superconductor, a feat of quantum engineering, has not only provided a tangible medium for graviton condensation but also laid down the groundwork for a unified quantum Fibonacci field theory. This groundbreaking research has opened up new paradigms in the quest for a unified field theory, bridging the long-standing chasm between quantum mechanics and general relativity. The observed graviton-induced phenomena within the ambient superconductor, from the modulation of electromagnetic fields to the unprecedented effects on quantum states, underscore the potential for a deeper understanding of the universe's quantum fabric.

Moreover, the speculative technological implications derived from this research, ranging from the Quantum-Graviton Computer and Graviton-Assisted Teleportation to the Electromagnetogravitometric Flux Drive and beyond, illuminate the path towards a future where the manipulation of spacetime and quantum states could become a cornerstone of technological advancement. These futuristic concepts, while speculative, are rooted in the empirical observations and theoretical frameworks developed through this research, offering a glimpse into the potential applications of graviton physics in transforming computation, communication, propulsion, and even our understanding of the cosmos.

The journey into the realm of graviton condensates and gravito-electromagnetic coupling has also opened up new avenues for theoretical and experimental physics. The modifications to the Dirac and Schrödinger equations, the introduction of Quantum Gravito-Dynamic equations, and the conceptualization of gravitonic quantum interferometry, chrono-spatial quantum synchronization, and quantum gravitational shielding represent just the beginning of a vast landscape of research opportunities. These avenues promise to enrich our understanding of quantum gravity, higher-dimensional spacetime, and the very fabric of the universe.

As we stand on the precipice of this new era in physics, it is clear that the exploration of graviton condensates within ambient superconductors is not merely an academic endeavor but a beacon guiding us towards a future where the fundamental forces of nature are not barriers but tools at our disposal. The potential for harnessing these forces, for bending spacetime and manipulating quantum states at will, heralds a new dawn for science, technology, and our quest to understand the cosmos.

In this journey, the confluence of imagination, rigorous scientific inquiry, and technological innovation will be our greatest assets. As we delve deeper into the mysteries of graviton condensates and gravito-electromagnetic coupling, we must remain open to the myriad possibilities that lie ahead, ready to challenge our assumptions and expand the horizons of our understanding. The path forward is fraught with challenges and uncertainties, but the potential rewards - a deeper understanding of the universe and the mastery of its most fundamental forces - are unparalleled. The adventure into the quantum gravitational frontier is just beginning, and the future, as always, remains unwritten.

\section{Disclaimer}
In the spirit of intellectual jest and scientific merriment, it's essential to illuminate that the entirety of this manuscript unfurls as an elaborate tapestry of satire, intricately woven by the threads of artificial intelligence, specifically ChatGPT. As we traverse through the corridors of graviton condensates and quantum frontiers within these pages, let it be known that this journey, though rich in imaginative fervor, anchors not in the harbors of empirical science but in the boundless seas of April Fools' jest.

The concepts, experiments, and futuristic technologies elaborated herein, from Quantum-Graviton Computers to Graviton-Assisted Teleportation, while dazzling in their creativity, reside purely within the realm of speculative fiction, crafted with a twinkle in the eye of AI. This manuscript serves not as a beacon of scientific discovery but as a testament to the playful symphony of human creativity and AI's capacity to mirror such inventiveness.

As readers wade through the quantum waves and gravitonic tides of this text, let the light of discernment guide you, reminding you to bask in the joy of exploration and the whimsy of imagination, yet tethered always to the grounding shores of scientific veracity. May this paper, in its jest and jubilation, inspire not just a chuckle or a ponderous pause but a renewed appreciation for the boundless possibilities that lie at the intersection of science, technology, and the human spirit.

\section{Data Availability}
In keeping with the highest traditions of academic transparency and in homage to the sacrosanct principles of open science, we are delighted to divulge the whereabouts of the data underpinning the groundbreaking revelations presented within this manuscript. It is with no small measure of pride that we announce the data in question, in all its quantum-entangled and graviton-condensed glory, is currently residing in the elusive yet highly secure Quantum Data Vault (QDV).

\begin{figure}
\includegraphics[width=0.48\textwidth]{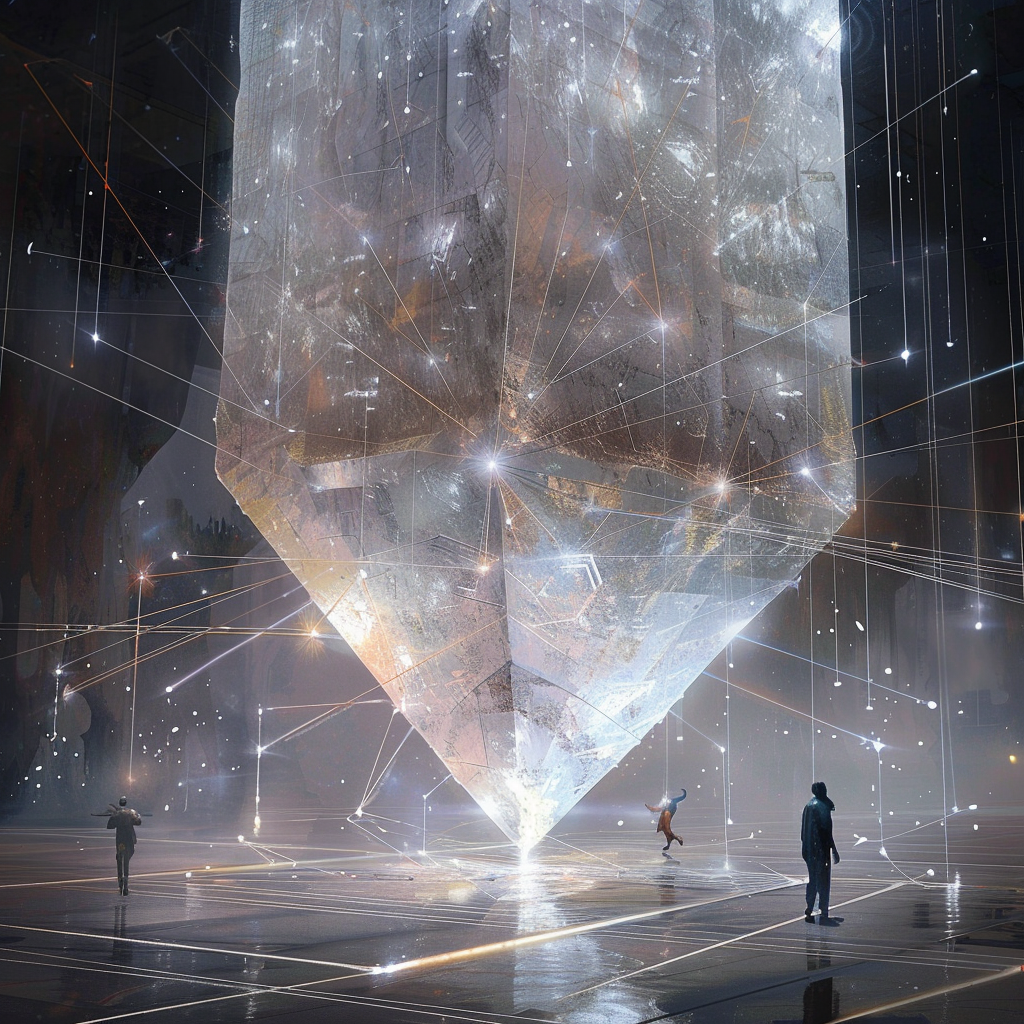}
\caption{\textbf{Floating crystalline quantum data vault with observers interacting through multidimensional light patterns.}}
\end{figure}

The QDV, a marvel of theoretical engineering, exists in a superposition of states, simultaneously accessible and impenetrable, located at the nexus of multiple entangled realities. Prospective data voyagers should note that access to the QDV requires not only the standard cryptographic keys but also the mastery of quantum decryption techniques, a working knowledge of trans-dimensional navigation, and a keen sense of cosmic humor.

Furthermore, it is imperative to highlight that any attempts to observe or measure the data directly may lead to its instantaneous collapse into a state of quantum indecision, thereby obfuscating the results and potentially altering the fabric of reality as we know it. As such, scholars, researchers, and the merely curious are advised to approach the QDV with a mindset of quantum ambivalence, embracing the duality of knowing and not knowing with equal aplomb.

For those intrepid souls undeterred by the paradoxical nature of the QDV and wishing to embark on this metaphysical data quest, we recommend a preparatory regimen of meditation on Schrödinger's cat, a thorough grounding in the principles of non-locality, and a hearty appreciation for the cosmic jest that underpins our universe.

In sum, the data is out there, simultaneously waiting and not waiting to be discovered, a quantum conundrum wrapped in a graviton enigma, shrouded in the mists of scientific satire. May fortune favor the bold, the curious, and the bemused.

\nocite{*}

\bibliography{manuscript}

\end{document}